\documentclass[draft]{aa}  
\usepackage{graphicx}
\usepackage{txfonts}

\usepackage{natbib}
\input{psfig}
\newcommand{\gtsim}{\protect\raisebox{-0.5ex}{$\:\stackrel{\textstyle >}
        {\sim}\:$}}

\begin{document}

    \title{Angular momentum conservation and torsional oscillations in the Sun and solar-like stars}

%    \title{On the source of the torsional oscillations in the Sun and in solar-like stars} 

    \titlerunning{Torsional oscillations in the Sun and stars}

%   \subtitle{}

   \author{Antonino~F.~Lanza}

   \offprints{A. F. Lanza}

   \institute{INAF-Osservatorio Astrofisico di Catania, Via S. Sofia, 78 -- 95123 Catania, Italy \\ 
              \email{nuccio.lanza@oact.inaf.it}    
             }

   \date{Received ... ; accepted ... }

    \abstract
{The solar torsional oscillations, i.e., the perturbations of the angular 
velocity of rotation associated with the eleven-year activity cycle, are a manifestation of the 
interaction among the interior magnetic fields, amplified and modulated by the 
solar dynamo, and rotation,  meridional flow and  turbulent thermal transport. 
Therefore, they can be used, at least in principle, to put constraints on that interaction. Similar
phenomena are expected to be observed in solar-like stars and can be 
modelled to shed light on analogous interactions in different environments.}
{The source of the torsional oscillations is investigated by means 
of a model for the angular momentum transport within the convection zone.}
{A description of the torsional oscillations is introduced, based on an analytical 
solution of the angular momentum equation in the mean-field approach. It provides information on the intensity and location of the 
torques producing the redistribution of the angular momentum  within the convection
 zone of the Sun along the activity cycle. The method can be extended to solar-like stars 
for which some information on the time-dependence of the differential rotation is becoming available.}
{ { Illustrative applications to the Sun and solar-like stars are presented.
Under the hypothesis that the solar torsional oscillations are due to the mean-field Lorentz force, 
an amplitude of the Maxwell stresses $|B_{\rm r}B_{\phi}| \gtsim 8 \times 10^{3}$ G$^{2}$ 
at a depth of $\sim 0.85$ $R_{\odot}$ at low latitude is estimated. Moreover, 
the phase relationship between $B_{\rm r}$ and $B_{\phi}$ can be estimated, suggesting that
$B_{\rm r} B_{\phi} > 0$ below $\sim 0.85$ $R_{\odot}$ and $B_{\rm r} B_{\phi} < 0$ above. }
}{ \rm Such preliminary results show the capability of the proposed approach to constrain the amplitude, phase and location of the perturbations leading to the observed torsional oscillations.}


   \maketitle
%
%________________________________________________________________

\section{Introduction}

Doppler measurements of the surface rotation of the Sun show 
bands of faster and slower zonal flows that appear at midlatitudes and migrate toward the 
equator with the period of the eleven-year cycle, accompanying the bands of sunspot activity. 
The amplitude of such velocity perturbations, called torsional oscillations, is of $\sim 5$ 
m s$^{-1}$ and  faster rotation is observed on the side equatorward of 
the sunspot belt \citep{HowardLabonte80}. Helioseismology has revealed that the torsional
oscillations are not at all a superficial phenomenon but involve much of the 
convection zone, as shown, for example, by \citet{Howeetal00}, \citet{Vorontsovetal02}, \citet{BasuAntia03}
and more recently by \citet{Howeetal05,Howeetal06}. The amplitude of the
angular velocity variation is $\delta \Omega/2\pi \sim 0.5-1$ nHz at least down to  10\%$-$15\% of the solar radius, 
although the precise depth of penetration
of the oscillations is difficult to establish given the present uncertainties of
the inversion methods in the lower half of the solar convection zone \citep[e.g., ][]{Howeetal06}.
In addition to such a low-latitude branch of the torsional oscillations, helioseismic studies have
detected the presence of a high-latitude branch (above $\sim 60^{\circ}$ latitude) that propagates poleward
and the amplitude of which is about $ \delta \Omega /2\pi \sim 1-2 $ nHz \citep[see also, e.g., ][]{Toomreetal00,BasuAntia01}. 
Such a branch seems to propagate almost all the way down to the base of the convection zone. 

A general description of the perturbation of the angular velocity of the torsional oscillations, given the present
accuracy of the observations, is provided by the simple formula \citep[cf., e.g.,][]{Vorontsovetal02,Howeetal05}:
\begin{eqnarray}
\omega (r, \theta) = A^{(c)} (r, \theta) \cos (\sigma t) + A^{(s)} (r, \theta) \sin (\sigma t) =  \nonumber \\
 = A(r, \theta) \sin[\sigma t + \phi(r, \theta)], 
\label{torsexp}
\end{eqnarray}
where $r$ is the distance from the centre of the Sun, $\theta$ the colatitude measured from the North pole,
 $\sigma$ the frequency of the
eleven-year cycle, $t$ the time, and the amplitude functions $A^{(c)} \equiv A \sin \phi$, $A^{(s)} \equiv A \cos \phi$
depend on the amplitude $A$ and the initial phase $\phi$. 
Moreover, the velocity perturbation is symmetric with respect to the equator:
\begin{equation}
\omega( r, \theta) = \omega (r, \pi - \theta). 
\label{eqsymm}
\end{equation}

Several models have been proposed to interpret the torsional oscillations beginning with the
pioneering work by \citet{Schussler81} and \citet{Yoshimura81} who considered  the Lorentz
force associated with the magnetic fields in the activity belts as the cause of the velocity perturbations
observed in the solar photosphere. Later models, based on the effects of the Lorentz force on the
turbulent Reynolds stresses, were proposed, by, e.g., \citet{Kukeretal96} and \citet{Kichatinovetal99}, 
following an original suggestion by \citet{RudigerKichatinov90}. 
\citet{Spruit03} proposed that the low-latitude branch of the torsional oscillations is a 
geostrophic flow driven by temperature variations due to the enhanced radiative losses in
the active region belts.

More recent works by \citet{Covasetal04,Covasetal05} present models based on the simultaneous solution of
non-linear mean-field dynamo equations and the azimuthal component of the Navier-Stokes
equation with a uniform turbulent viscosity. They reproduce the gross features of the 
torsional oscillations and of the solar activity cycle with an appropriate tuning of the free
parameters. \citet{Rempel06,Rempel07} considers the role of the meridional component of the
Navier-Stokes equation in mean-field models and finds that the perturbation of the meridional flow
cannot be neglected in the interpretation of the torsional oscillations. His models suggest that
the low-latitude branch of the torsional oscillations cannot be explained solely by the effect of 
the mean-field Lorentz force, but that thermal perturbations in the active region belt and
in the bulk of the convection zone do play an active role, as proposed by \citet{Spruit03}. 
 
In the present study, the angular momentum conservation  is
considered and the relevant equation in the mean-field approximation is solved analytically
for the case of a turbulent viscosity that depends on the radial co-ordinate. A general solution is derived independently of
any specific dynamo model, allowing us to put constraints on the localization of the torques
producing the torsional oscillations. An illustrative application of 
the proposed methods is presented using the available data.

The observations of young solar-like stars by means of tomographic techniques based on high-resolution
spectroscopy have recently provided evidence for time variation of their surface 
differential rotation \citep[see, e. g., ][]{Donatietal03, Jeffersetal07}. \citet{Lanza06}
has recently shown how such variations can be related to the 
intensity of the magnetic torque produced by a non-linear dynamo in their convective
envelopes, in the case of rapidly rotating stars for which the Taylor-Proudman theorem applies. 
In the near future, the possibility of
measuring the time variation of the rotational splittings of p-mode oscillations in solar-like stars
may provide us with  information on the changes of their internal rotation, although 
with limited spatial resolution. In the present study,
we extend the considerations of \citet{Lanza06} to the case of a generic internal rotation profile,
not necessarily verifying the Taylor-Proudman theorem, to obtain  hints on the amplitude of 
the torque leading to the rotation change. 

\section{The model}

\subsection{Hypotheses and basic equations}

We consider an inertial reference frame with the origin in the 
barycentre of the Sun and the $z$-axis in the direction of
the rotation axis. A spherical polar co-ordinate system $(r, \theta, \varphi)$ is adopted,
where $r$ is the distance from the origin, $\theta$ the co-latitude measured from the
North pole and $\varphi$ the azimuthal angle. We assume that all the variables
are independent of $\varphi$ and that the solar density stratification is spherically symmetric.

The equation for the angular momentum conservation in the mean-field approach reads
\citep[e.g., ][]{Rudiger89,RudigerHollerbach04}:
\begin{equation}
\frac{\partial}{\partial t} (\rho r^{2} \sin^{2} \theta \, \Omega) + \nabla \cdot {\vec \Theta} = 0,  
\label{angmomeq}
\end{equation}
where $\rho(r)$ is the density, $\Omega (r, \theta, t)$ the angular velocity and ${\vec \Theta}$ 
the angular momentum flux vector given by:
\begin{eqnarray}
\lefteqn{{\vec \Theta} = (\rho r^{2} \sin^{2} \theta \, \Omega) {\vec u}_{\rm (m)} +} \nonumber \\
 & & + r \sin \theta \langle \rho {\vec u}^{\prime} u_{\varphi}^{\prime} \rangle  
- \frac{r \sin \theta}{\tilde{\mu}} ({\vec B} B_{\varphi} + \langle {\vec B}^{\prime} B_{\varphi}^{\prime} \rangle),
\end{eqnarray}
where ${\vec u}_{\rm (m)}$ is the meridional circulation, ${\vec u}^{\prime}$ the fluctuating velocity field, $\tilde{\mu}$
the magnetic permeability, ${\vec B}$ the mean magnetic field and ${\vec B}^{\prime}$ the fluctuating magnetic
field; angular brackets indicate the Reynolds average defining the mean-field quantities. The Reynolds
stresses can be written as: 
\begin{equation}
\langle \rho u^{\prime}_{i} u^{\prime}_{j} \rangle = -\eta_{\rm t} \left( \frac{\partial u_{i}}{\partial x_{j}} +
\frac{\partial u_{j}}{\partial x_{i}} \right) + \Lambda_{ij},  
\end{equation}
where ${\vec u}$ is the mean flow field, $\eta_{\rm t}(r)$ is the turbulent viscosity, assumed to be a scalar function of 
$r$ only, and $\Lambda_{ij}$ indicates the non-diffusive part of the Reynolds stresses due to the velocity
correlations in a rotating star \citep[see ][ for details]{Rudiger89,RudigerHollerbach04}. The conservation of the total
angular momentum of the convection zone implies:
\begin{equation}
\Theta_{r} = 0 \mbox{ for $r=r_{\rm b}, \, R_{\odot}$,}
\label{bcond}
\end{equation}
where $r_{\rm b}$ is the radius at the lower boundary of the convection zone and $R_{\odot}$ is the radius of the Sun. 

The equation for the conservation of the angular momentum can be recast in the form: 
\begin{eqnarray}
\lefteqn{\frac{\partial \Omega}{\partial t} - \frac{1}{\rho r^{4}} \frac{\partial}{\partial r} 
\left(  r^{4} \eta_{t} \frac{\partial \Omega}{\partial r} \right) +}   \nonumber \\
 &  & -\frac{\eta_{t}}{\rho r^{2}}
\frac{1}{(1- \mu^{2})} \frac{\partial}{\partial \mu} \left[ (1 - \mu^{2})^{2} \frac{\partial \Omega}{\partial \mu}\right] = S,  
\label{angvel}
\end{eqnarray} 
where $\mu \equiv \cos \theta$ and the source term $S$ is given by:
\begin{equation}
S = - \frac{\nabla \cdot {\vec \tau}}{\rho r^{2} (1- \mu^{2})}, 
\label{sourceangvel}
\end{equation}
and ${\vec \tau}$ is a vector whose components are: 
\begin{equation}
\tau_{i} = r \sin \theta \left[ \Lambda_{i \varphi} - \frac{1}{\tilde{\mu}} \left( B_{i} B_{\varphi} + 
 \langle B_{i}^{\prime} B_{\varphi}^{\prime} \rangle \right) \right] + \rho r^{2} \sin^{2} \theta \Omega  u_{\rm (m) i}. 
\label{tau}
\end{equation}
The boundary conditions given by Eq.~(\ref{bcond}) can be written as:
\begin{equation}
\frac{\partial \Omega}{\partial r} = 0 \mbox{ for $r=r_{\rm b}, \, R_{\odot}$,}
\label{bcond1}
\end{equation}
{
when we assume $\tau_{\rm r} = 0$ at the surface. 
Note that helioseismic measurements indicate the presence of a subsurface shear layer
with $\frac{\partial \Omega}{\partial r} < 0$ at low latitudes \citep{CorbardThompson05}, but we prefer to adopt the stress-free
boundary condition (\ref{bcond1}) at the surface to ensure the conservation of the 
total angular momentum of the convection zone in our model. } 

The solar angular velocity can be split into a time-independent component
$\Omega_{0}$ and a time-dependent component $\omega$, i.e., the torsional
oscillations: 
\begin{equation}
\Omega (r, \mu, t) = \Omega_{0} (r, \mu) + \omega (r, \mu, t).  
\end{equation}
The equation for the torsional oscillations thus becomes: 
\begin{eqnarray}
\lefteqn{\frac{\partial \omega}{\partial t} - \frac{1}{\rho r^{4}} \frac{\partial}{\partial r} 
\left(  r^{4} \eta_{t} \frac{\partial \omega}{\partial r} \right) +}   \nonumber \\
 &  & -\frac{\eta_{t}}{\rho r^{2}}
\frac{1}{(1- \mu^{2})} \frac{\partial}{\partial \mu} \left[ (1 - \mu^{2})^{2} \frac{\partial \omega}{\partial \mu}\right] = S_{1},  
\label{angvelp}
\end{eqnarray} 
where the perturbation of the source term is:
\begin{equation}
S_{1} =  - \frac{\nabla \cdot {\vec \tau}_{1}}{\rho r^{2} (1- \mu^{2})},
\label{s1}
\end{equation}
with 
\begin{equation}
\tau_{1 i} = r \sin \theta \left[ \Lambda_{i \varphi}^{\rm (p)} - \frac{1}{\tilde{\mu}} \left( B_{i} B_{\varphi} 
+ \langle B_{i}^{\prime} B_{\varphi}^{\prime} \rangle \right) \right] + \rho r^{2} \sin^{2} \theta \bar{\Omega}_{0} u^{\rm (p)}_{\rm (m) i}, 
\label{tau1}
\end{equation}
where $\Lambda_{i \varphi}^{\rm (p)}$ and $u^{\rm (p)}_{\rm (m) i}$ are the time-dependent perturbations of 
the non-diffusive Reynolds stresses and of the meridional circulation, respectively, and $\bar{\Omega}_{0}$ is the 
average of the solar angular velocity over the convection zone. Note that  the Maxwell stresses 
appear in Eq.~(\ref{tau1}), but not in the corresponding equation for 
$\Omega_{0}$ because the solar magnetic field has no time-independent component. 
Moreover, in deriving Eq.~(\ref{tau1}), the variation of the angular velocity 
over the convection zone has been neglected in the term containing the perturbation of the meridional
circulation since $|\omega| \ll \Omega_{0}$ \citep[cf., e.g., ][]{Rudiger89,Rempel07}. Eq.~(\ref{angvelp})
must be solved together with the boundary conditions:
\begin{equation}
\frac{\partial \omega}{\partial r} = 0 \mbox{ for $r=r_{b}, \, R_{\odot}$.}
\label{bcond2}
\end{equation}

\subsection{Solution of the angular momentum equation }
\label{general_sol}

The general solution of Eq.~(\ref{angvelp}) with the boundary conditions (\ref{bcond2}) can be obtained by the method of
separation of the variables and expressed as a series of the form \citep[cf., e.g., ][ and references therein]{Lanza06b}:
\begin{equation}
\omega (r, \mu , t) = \sum_{n = 0, 2, 4, ...}^{\infty} \sum_{k=0}^{\infty} \alpha_{nk} (t) \zeta_{nk} (r) P_{n}^{(1,1)} (\mu),
\label{omegadevel}
\end{equation}
where $\alpha_{nk}(t)$ and $\zeta_{nk}(r)$ are functions  that will be specified below 
and $P_{n}^{(1, 1)}(\mu)$ are Jacobian polynomials, i.e., the finite solutions of the equation: 
\begin{equation}
\frac{d}{d \mu} \left[ (1- \mu^{2})^{2} \frac{d P_{n}^{(1, 1)}}{d \mu }\right] + 
n(n+3) (1 - \mu^{2}) P_{n}^{(1, 1)} = 0, 
\label{jacobian}
\end{equation} 
in the interval $ -1 \leq \mu \leq 1$ including its ends \citep[cf., e.g., ][]{Smirnov64a}. 
The Jacobian polynomials form a complete and orthogonal set in the interval $[-1, 1]$ with 
respect to the weight function $(1-\mu^{2})$. Only the polynomials of even degree appear in Eq.~(\ref{omegadevel})
because the angular velocity perturbation is symmetric with respect to the equator (see Eq.~\ref{eqsymm}). 
For $n \gg 1$ the asymptotic expression of the Jacobian polynomials is \citep[see, e.g., ][]{GradshteynRizhik94}:
\begin{equation}
P_{n}^{(1,1)} (\cos \theta) =
 \frac{\cos \left[ (n + \frac{3}{2}) \theta 
- \frac{3\pi}{4}\right]}{\sqrt{\pi n} \left[\sin \left(\frac{\theta}{2}\right) \cos \left(\frac{\theta}{2}\right)\right]^{3/2}} 
+ O( n^{-\frac{3}{2}}) 
\label{jacobi_asymp}
\end{equation}

The functions $\zeta_{nk}$ are the solutions of the Sturm-Liouville 
problem defined in the interval $r_{b} \leq r \leq R_{\odot}$ by the equation:
\begin{equation}
 \frac{d}{dr} \left( r^{4} \eta_{\rm t} \frac{d \zeta_{nk}}{dr} \right) - n(n+3) r^{2} \eta_{\rm t}
\zeta_{nk} + \lambda_{nk} \rho r^{4} \zeta_{nk} = 0 
\label{radialpart}
\end{equation}
with the boundary conditions (following from Eq.~\ref{bcond2}): 
\begin{equation}
\frac{d \zeta_{nk}}{dr} = 0 \mbox{ at $r=r_{b}, R_{\odot}$.}
\label{bc1}
\end{equation} 
We shall consider normalized eigenfunctions, i.e.: $\int_{r_{\rm b}}^{R_{\odot}} \rho r^{4} \zeta_{nk}^{2} dr = 1$. 
The eigenfunctions $\zeta_{nk}$ for a fixed $n$, form a complete and orthonormal set 
in the interval $[r_{\rm b}, R_{\odot}]$ with respect to the weight function $\rho r^{4}$ that does not depend on $n$. We recall
from the theory of the Sturm-Liouville problem that the eigenvalues verify the inequality:
$ \lambda_{n 0} < \lambda_{n 1} <...\lambda_{n k}< \lambda_{n k+1}< ... $ and that the
eigenfunction $\zeta_{n k}$ has $k$ nodes in the interval $[r_{b}, R_{\odot}]$ for each $n$.  For $n=0$, the first eigenvalue
corresponding to the eigenfunction $\zeta_{00}$ is zero and the eigenfunction vanishes at all points in 
$[r_{b}, R_{\odot}]$, as it is evident by integrating both sides of Eq. (\ref{radialpart}) in the same interval, applying the 
boundary conditions (\ref{bc1}) and considering that $\zeta_{00}$ has no nodes. For $n >0$, all the eigenvalues $\lambda_{n0}$ are
positive, as can be derived by integrating both sides of Eq. (\ref{radialpart}) in the  interval $[r_{b}, R_{\odot}]$,
applying the boundary conditions (\ref{bc1}) and considering that $\zeta_{n0}$ has no nodes. In view of the inequality
given above, all the eigenvalues $\lambda_{nk}$ are then positive for $ n\geq 0$. Moreover, it is possible to prove
that  $\lambda_{n^{\prime} k} \geq \lambda_{n k}$ if $n^{\prime} > n$ and that \citep[see ][]{Smirnov64b}:
\begin{equation}
\frac{\left(\frac{k^{2} \pi^{2}}{l^{2}}\right) p + n(n+3) q}{M} \leq \lambda_{nk} 
\leq \frac{\left(\frac{k^{2} \pi^{2}}{l^{2}} \right) P + n(n+3) Q}{m}, 
\label{eigen_ineq}
\end{equation}
where $P$, $Q$ and $M$ are the maximum values of the functions $r^{4}\eta_{\rm t}$, $r^{2}\eta_{\rm t}$ and $\rho r^{4}$ in the 
interval $[r_{b}, R_{\odot}]$, respectively, and $p$, $q$ and $m$ their minimum values in the same interval, respectively; and 
$l \equiv \int_{r_{b}}^{R_{\odot}} \sqrt{\rho / \eta_{\rm t}} dr$. For  $k \gg 1$ the asymptotic expression for
the eigenfunction $\zeta_{nk}$ is \citep[cf., e.g., ][]{MorseFeshbach53}:
\begin{equation}
\zeta_{nk}(r) \simeq \left( \rho \eta_{\rm t}\right)^{-\frac{1}{4}} r^{-2} \cos \left( \sqrt{\lambda_{nk}}
 \int_{r_{b}}^{r} \sqrt{\rho/\eta_{\rm t}} d r^{\prime} \right). 
\label{zeta_asymp}
\end{equation}  
 
The time dependence of the solution (\ref{omegadevel}) is specified by the functions $\alpha_{nk}$ that are given by:
\begin{equation}
\frac{d \alpha_{nk}}{dt} + \lambda_{nk} \alpha_{nk}(t) = \beta_{nk} (t),
\label{alphaeq}
\end{equation}
where the functions $\beta_{nk}$ appear in the development of the perturbation term $S_{1}$:
\begin{equation}
S_{1}(r, \mu, t) = \sum_{n} \sum_{k} \beta_{nk} (t) \zeta_{nk} (r) P_{n}^{(1,1)}(\mu),
\label{s1devel}
\end{equation}
and are given by \citep[cf. ][]{Lanza06b}:
\begin{eqnarray}
\lefteqn{\beta_{nk} = \frac{(2n+3) (n+2)}{8(n+1)} \, \times} \nonumber \\ 
& & \times \int_{r_{\rm b}}^{R_{\odot}} \int_{-1}^{1} \rho r^{4} (1-\mu^{2}) S_{1}(r, \mu, t) \zeta_{nk}(r) 
P_{n}^{(1,1)} (\mu) dr d\mu, 
\label{betaeq}
\end{eqnarray}
The solution of Eq.~(\ref{alphaeq}) is:
\begin{equation}
\alpha_{nk}(t) = \alpha_{nk} (0) + \exp(-\lambda_{nk} t) \int_{0}^{t} \beta_{nk}(t^{\prime}) \exp(\lambda_{nk} t^{\prime})
 d t^{\prime}. 
\label{alphabeta}
\end{equation}
which allows us to specify the general solution of Eq.~(\ref{angvelp}) with the boundary conditions 
(\ref{bcond2}) when the perturbation term $S_{1}(r, \mu, t)$ and the 
initial conditions are given.  

\subsection{Solution for the solar torsional oscillations}
\label{sol_method}

To find the solution appropriate to the solar torsional oscillations as specified by Eqs.~(\ref{torsexp}) and (\ref{eqsymm}),
it is useful to derive an alternative expression for the functions $\beta_{nk}$ as follows. Substituting Eq.~(\ref{s1})
into Eq.~(\ref{betaeq}) and taking into account that the element of volume is $dV = - 2 \pi r^{2} dr d\mu$, 
the r.h.s. of Eq.~(\ref{betaeq}) can be recast in the form of a volume integral extended to the solar convection zone:
\begin{equation}
\beta_{nk} = F_{n} \int_{V} (\nabla \cdot {\vec \tau}_{1}) \zeta_{nk} P_{n}^{(1,1)} dV,
\label{betaeq1} 
\end{equation}
where:
\begin{equation}
F_{n} \equiv \frac{(2n+3) (n+2)}{ 16\pi (n+1)}
\end{equation}
is a factor coming from the normalization of the Jacobian polynomials. 
It is possible to simplify further Eq.~(\ref{betaeq1}) by considering the identity:
\begin{equation}
\nabla \cdot [\zeta_{nk} P_{n}^{(1,1)} {\vec \tau}_{1}] = \zeta_{nk} P_{n}^{(1,1)} (\nabla \cdot {\vec \tau}_{1}) +
{\vec \tau}_{1} \cdot \nabla[\zeta_{nk} P_{n}^{(1,1)}].
\label{divtheor}
\end{equation}
Integrating both sides of  (\ref{divtheor}) over the volume of the convection zone 
and considering that the integral of the l.h.s. vanishes 
thanks to the Gauss's theorem and the condition that the radial component of the stresses $\tau_{1 r}$ is zero
on the boundaries, we find:
\begin{equation}
\beta_{nk} = - F_{n} \int_{V} {\vec \tau}_{1} \cdot \nabla[\zeta_{nk} P_{n}^{(1,1)}] dV. 
\label{betaeq2} 
\end{equation}

The time dependence of the solar torsional oscillations specified in Eq.~(\ref{torsexp}) suggests to consider a 
similar dependence for the perturbation term:
\begin{equation}
{\vec \tau}_{1} (r, \mu, t) = {\vec \tau}_{1}^{(c)}(r, \mu) \cos (\sigma t) + {\vec \tau}_{1}^{(s)}(r, \mu) \sin (\sigma t),
\label{tautimedep}
\end{equation}
from which a similar expression for the $\beta_{nk}$ follows by substitution into Eq.~(\ref{betaeq1}).
If we put such an expression for $\beta_{nk}$ into 
Eq.~(\ref{alphabeta}) and perform the integrations with respect to the time, we find the stationary solution:
\begin{eqnarray}
\lefteqn{\alpha_{nk} (t) =  \frac{F_{n}}{\lambda_{nk}^{2} + \sigma^{2}} \times}  \nonumber \\
& &  
\times \left\{ \left[ \int_{V} \zeta_{nk} P_{n}^{(1,1)} 
\left( \lambda_{nk} \nabla \cdot {\vec \tau}_{1}^{(c)} - \sigma \nabla \cdot {\vec \tau}_{1}^{(s)} \right) 
dV \right] \cos (\sigma t) + \right.  \nonumber \\
& & \left. + \left[ \int_{V} \zeta_{nk} P_{n}^{(1,1)} 
\left( \lambda_{nk} \nabla \cdot {\vec \tau}_{1}^{(s)} + \sigma \nabla \cdot {\vec \tau}_{1}^{(c)} \right) dV \right] \sin (\sigma t) \right\}  
\label{alphafin}
\end{eqnarray}
This expression can be substituted into Eq.~(\ref{omegadevel}) to give the angular velocity
perturbation. It can be written in the form of Eq.~(\ref{torsexp}) with:
\begin{eqnarray}
A^{(c)} (r, \mu) =\!\!  \int_{V} \left[ G_{1}(r, \mu , r^{\prime}, \mu^{\prime}) \nabla \cdot  {\vec \tau}_{1}^{(c)} -
G_{2}(r, \mu , r^{\prime}, \mu^{\prime}) \nabla \cdot  {\vec \tau}_{1}^{(s)} \right] d V^{\prime},  \nonumber \\
A^{(s)} (r, \mu) = \!\! \int_{V} \left[ G_{1}(r, \mu , r^{\prime}, \mu^{\prime}) \nabla \cdot  {\vec \tau}_{1}^{(s)} + 
G_{2}(r, \mu , r^{\prime}, \mu^{\prime}) \nabla \cdot  {\vec \tau}_{1}^{(c)} \right] d V^{\prime}, \nonumber \\
 & & 
\label{A_sol}
\end{eqnarray}
where the symbol $d V^{\prime}$ means that the volume integration is to be performed with respect to the variables
$r^{\prime}$ and $\mu^{\prime}$,  and the functions $G_{1}$ and $G_{2}$ are Green functions defined as: 
\begin{eqnarray}
G_{1}(r, \mu , r^{\prime}, \mu^{\prime}) = \sum_{n}^{\infty} F_{n} \sum_{k}^{\infty} 
\frac{\lambda_{nk}}{\lambda_{nk}^{2} + \sigma^{2}} \times \nonumber \\
\times \zeta_{nk}(r) P_{n}^{(1,1)}(\mu)  
\zeta_{nk}(r^{\prime}) P_{n}^{(1,1)}(\mu^{\prime}) \nonumber \\
G_{2}(r, \mu , r^{\prime}, \mu^{\prime}) = \sum_{n}^{\infty} F_{n} \sum_{k}^{\infty} 
\frac{\sigma}{\lambda_{nk}^{2} + \sigma^{2}} \times \nonumber \\ 
\times \zeta_{nk}(r) P_{n}^{(1,1)}(\mu) 
\zeta_{nk}(r^{\prime}) P_{n}^{(1,1)}(\mu^{\prime}). 
\label{green}
\end{eqnarray}
The Green functions are continuous with respect to the arguments $r$, $\mu$, $r^{\prime}$, $\mu^{\prime}$, but their
partial derivatives with respect to $r^{\prime}$, $\mu^{\prime}$ have discontinuities of the first kind in the 
points where $r=r^{\prime}$ or $\mu=\mu^{\prime}$. The convergence of the series  in Eqs. (\ref{green}), 
here used to represent the Green functions, is assured by the general theory of the Green function 
 \citep[see, e.g., ][]{Smirnov64b} and is also proven in Appendix A.

Eqs.~(\ref{A_sol}) can be used to compute the angular velocity perturbation when ${\vec \tau}_{1}$ is known.
Note that in the case in which the  Lorentz force due to the mean field and
the meridional flow are the only sources of angular momentum redistribution, Eq.~(\ref{tau1}) gives:
\begin{equation}
\nabla \cdot {\vec \tau}_{1} = - \frac{1}{\tilde{\mu}}{\vec B}_{\rm p} \cdot \nabla ( r \sin \theta B_{\phi}) + 
2 \rho \bar{\Omega}_{0}   (r \sin \theta) u_{\rm (m) s}, 
\label{expdivtau}
\end{equation}
where ${\vec B}_{\rm p}$ is the mean poloidal magnetic field and $u_{\rm (m) s}$ the component of the meridional flow in the direction orthogonal to the rotation axis. To obtain Eq.~(\ref{expdivtau}) we 
made use of the solenoidal nature of the mean poloidal field and of the continuity equation for the meridional flow.

\subsection{Localization of the source of the torsional oscillations in the solar convection zone}
\label{loc_method}

The results derived above allow us to introduce 
methods to localize the torques producing the torsional oscillations in the convection
zone. Suppose that the observations provide us with the functions $A^{(c)}(r, \mu)$ and
$A^{(s)}(r, \mu)$ appearing in Eq.~(\ref{torsexp}). The functions $\alpha_{nk}(t)$ can be 
written as: 
\begin{equation}
\alpha_{nk} = a_{nk}^{(c)} \cos (\sigma t) + a_{nk}^{(s)} \sin (\sigma t), 
\end{equation}
where the constants $a_{nk}^{(c,s)}$ are given by:
\begin{equation}
a_{nk}^{(c,s)} = 2\pi F_{n} \int_{r_{\rm b}}^{R_{\odot}} \int_{-1}^{1} A^{(c,s)}(r,\mu) \rho r^{4} (1-\mu^{2}) 
\zeta_{nk} P_{n}^{(1,1)} dr d\mu.
\label{acoeff}
\end{equation}
Similarly, we can write:
\begin{equation}
\beta_{nk}(t) =  b_{nk}^{(c)} \cos (\sigma t) + b_{nk}^{(s)} \sin (\sigma t),
\label{bdef}
\end{equation}
with the relationships:
\begin{eqnarray}
b_{nk}^{(c)} = \lambda_{nk} a_{nk}^{(c)} + \sigma a_{nk}^{(s)}, \nonumber \\
b_{nk}^{(s)} = \lambda_{nk} a_{nk}^{(s)} - \sigma a_{nk}^{(c)},  
\label{bcoeff}
\end{eqnarray}
that follow from Eq.~(\ref{alphaeq}). 

The divergence of ${\vec \tau}_{1}^{(c,s)}$ can be obtained from  
Eqs.~(\ref{s1}) and (\ref{s1devel}) as:
\begin{equation}
\nabla \cdot {\vec \tau}_{1}^{(c,s)} = - \rho r^{2} (1-\mu^{2}) \sum_{n} \sum_{k} b_{nk}^{(c, s)} 
\zeta_{nk} (r) P_{n}^{(1,1)} (\mu). 
\label{div_loc}
\end{equation}
Moreover, it is possible to construct a localized estimate of the perturbation term 
$ {\vec \tau}_{1}$ by considering a function
$f(r, \mu)$ the gradient of which is different from zero only within a given volume
$V_{f}$. It can be developed in series of the eigenfunctions in the form:
\begin{equation}
f(r, \mu) = \sum_{n} \sum_{k} c_{nk} \zeta_{nk} (r) P_{n}^{(1,1)} (\mu),  
\label{fexp}
\end{equation}
where the coefficient $c_{nk}$ are given by:
\begin{equation}
c_{nk} = 2\pi F_{n} \int_{r_{\rm b}}^{R_{\odot}} \int_{-1}^{1} f(r,\mu) \rho r^{4} (1-\mu^{2}) \zeta_{nk} P_{n}^{(1,1)} dr d\mu.
\end{equation}
Let us consider the equation: 
\begin{eqnarray}
{\int_{V_{\rm f}} \left( {\vec \tau}_{1}^{(c,s)} \cdot \nabla f \right) dV = }  & & \nonumber \\ 
= \int_{V_{\rm f}} \left( {\vec \tau}_{1}^{(c,s)} \cdot \sum_{n} 
\sum_{k} \nabla [c_{nk} \zeta_{nk} (r) P_{n}^{(1,1)} (\mu)] \right) dV, 
\label{eq42}
\end{eqnarray}
obtained by means of Eq.~(\ref{fexp}). 
Considering Eqs.~(\ref{betaeq2}) and (\ref{bdef}), Eq.~(\ref{eq42}) can be recast as:
\begin{eqnarray}
\int_{V_{\rm f}} \left( {\vec \tau}_{1}^{(c,s)} \cdot \nabla f \right) dV = - \sum_{n} \frac{1}{F_{n}} \sum_{k} c_{nk} b_{nk}^{(c,s)}.
\label{averagetau} 
\end{eqnarray}
Moreover, if we introduce the volume average of the modulus of the perturbation 
$|{\vec \tau}_{1}|$ with respect to the weight function $|\nabla f|$, i.e.:
\begin{equation}
\langle | {\vec \tau}_{1}^{(c,s)} | \rangle_{V_{\rm f}} \equiv \frac{\int_{V_{\rm f}} | {\vec \tau}_{1}^{(c,s)} | | \nabla f | dV}{  
\int_{V_{\rm f}} |\nabla f| dV},   
\end{equation}
then Eq.~(\ref{averagetau}) gives a lower limit for it in the form:
\begin{equation}
\langle | {\vec \tau}_{1}^{(c,s)} | \rangle_{V_{\rm f}} 
\geq \frac{\sum_{n} \frac{1}{F_{n}} \sum_{k} c_{nk} b_{nk}^{(c,s)}}{\int_{V_{\rm f}} |\nabla f| dV}.
\label{lower_tau1}
\end{equation}

The minimum dimensions of the volume $V_{\rm f}$ are set by the spatial resolution of the measurements of the angular velocity
variations. They depend on the accuracy of the rotational splitting
coefficients, the inversion technique 
and the position within the convection zone \citep[see, e.g., ][]{Schouetal98,Howeetal05}. 
The minimum order of the Jacobian polynomials $N_{\rm m}$ 
needed to reproduce an angular velocity variation with a latitudinal resolution $\Delta \theta$ is 
$N_{\rm m} \sim \frac{2\pi}{\Delta \theta}$. Similarly, the minimum order of the radial eigenfunctions $K_{\rm m}$ is set by the
radial resolution $\Delta r$ as $ K_{\rm m} \sim \frac{2(R-r_{\rm b})}{\Delta r}$. Therefore, it is possible to truncate the
series  in Eqs.~(\ref{averagetau}) and (\ref{lower_tau1}) to those upper limits
for $n$ and $k$ because the coefficients $c_{nk}$ will decrease
rapidly for $n> N_{\rm m}$ and $k > K_{\rm m}$ giving a negligible contribution to the sum. 

The statistical errors in the measurements of the angular velocity variations can be easily propagated through the 
linear equations (\ref{acoeff}), (\ref{bcoeff}), (\ref{div_loc}) and (\ref{averagetau}) to find the errors on the estimates of 
$\nabla \cdot {\vec \tau}_{1}^{(c,s)}$ or the average of ${\vec \tau}_{1}$. For instance, if we consider 
the standard deviations $\sigma_{i}$ of the data $d_{i}$, i.e., the rotational 
splittings or the splitting coefficients from which the internal rotation is derived, the standard deviation
$\sigma_{\rm I}$ of the integral in Eq. (\ref{averagetau}) is:
\begin{equation}
\sigma_{\rm I}^{2} \simeq  \sum_{n}  \frac{1}{F_{n}^{2}} \sum_{k} c_{nk}^{2} \lambda_{nk}^{2} \sum_{i} e_{ink}^{2} \sigma_{i}^{2},
\end{equation}
where 
\begin{equation}
e_{ink} \equiv 2 \pi F_{n} \int_{r_{\rm b}}^{R_{\odot}} \int_{-1}^{1} (1-\mu^{2}) 
\rho r^{4} c_{i} (r, \mu) \zeta_{nk} P_{n}^{(1,1)} dr d\mu, 
\end{equation}
and the functions $c_{i} (r, \mu)$ are the rotational inversion coefficients defined in Eq. (8) of \citet{Schouetal98}. 

Note that a constant relative error $\epsilon = \Delta \omega/ \omega$ in the measurements of 
$A^{(c,s)}$ leads to the same relative error in Eq.~(\ref{div_loc}) and in Eqs. (\ref{averagetau}) and (\ref{lower_tau1}), 
given the linear equations
that relate the corresponding quantities. As a matter of fact, there is also a systematic error in our inversion method 
related to the poor knowledge of the turbulent viscosity $\eta_{\rm t}(r)$ that determines the form of the radial eigenfunctions
$\zeta_{nk}$.  

\subsection{Kinetic energy variation and dissipation}

The variation of the kinetic energy of rotation associated with the torsional oscillations, averaged over the  
eleven-year cycle, can be computed after \citet{Lanza06b} and it is:
\begin{equation}
\langle \Delta {\cal T} \rangle_{\rm c}= \sum_{n} \sum_{k} \langle \Delta {\cal T}_{nk} \rangle_{\rm c},
\label{kvar}
\end{equation}
where
\begin{equation}
\langle \Delta {\cal T}_{nk} \rangle_{\rm c} = \frac{1}{4F_{n}} \left\{ [a_{nk}^{(c)}]^{2}  + [a_{nk}^{(s)}]^{2} \right\}.  
\end{equation}
The average dissipation rate of the kinetic energy of the torsional oscillations due to the turbulent 
viscosity is:
\begin{equation}
\langle \frac{d {\cal T}}{dt} \rangle_{\rm c} = -2 \sum_{n} \sum_{k} \lambda_{nk} \langle \Delta {\cal T}_{nk} \rangle_{\rm c}.
\label{kdiss}
\end{equation}

{
\subsection{Torsional oscillations due to mean-field Lorentz force}
\label{meanfieldeff}

Most models of the torsional oscillations assume that they are due to the Lorentz force
produced by the mean field as derived from dynamo models. Therefore, let us consider the case in which only the
mean-field Maxwell stresses contribute to the perturbation, i.e.:
\begin{equation}
{\vec \tau}_{1} = - \frac{1}{\tilde{\mu}} (r \sin \theta)  B_{\phi} \, {\vec B}_{\rm p}.
\end{equation}
If the mean radial $B_{\rm r}$ and toroidal fields $B_{\phi}$ are given by:
\begin{eqnarray}
B_{\rm r} & = & B_{\rm 0 r} \cos (\frac{1}{2} \sigma t), \nonumber \\
B_{\phi} &  = & B_{0 \phi} \cos (\frac{1}{2} \sigma t + \Pi_{\rm r}), 
\end{eqnarray}
where $\Pi_{\rm r}$ is the phase lag between the two field components, 
the components of $\tau_{1r}$ in Eq.~(\ref{tautimedep}) are:
\begin{eqnarray}
\tau_{1r}^{(c)} & = & -\frac{1}{2} \cos \Pi_{\rm r} \, \frac{r \sin \theta}{\tilde{\mu}} \,  B_{\rm 0 r} B_{0 \phi}, \nonumber \\
\tau_{1r}^{(s)} & = & \frac{1}{2} \sin \Pi_{\rm r} \, \frac{r \sin \theta}{\tilde{\mu}} \, B_{\rm 0 r} B_{0 \phi}.
\label{brbphi}
\end{eqnarray}
An estimate of $\tau_{1r}^{(c)}$ and $\tau_{1r}^{(s)}$ can be obtained from the method outlined 
in Sect.~\ref{loc_method}, considering a localization function $f(r)$ that depends only on the radial
co-ordinate. Specifically, Eq.~(\ref{averagetau}) can be used to compute a volume average of
$\tau_{1r}^{(c)}$ and $\tau_{1r}^{(s)}$ from which the average stress amplitude $|B_{\rm r} B_{\phi} | $ and 
phase lag $\Pi_{\rm r}$ can be determined. Analogous considerations are valid for the meridional component
of the mean field $B_{\theta}$ and $B_{\phi}$. Adopting a localization function $f(\theta)$ depending only on $\theta$,
it is possible to estimate $|B_{\theta} B_{\phi}|$ and the phase lag $\Pi_{\theta}$ between $B_{\theta}$ and $B_{\phi}$. 
Such results are important to constrain mean-field dynamo models of the solar cycle, as discussed by, e. g.,
\citet{SchlichenStix95} (see also Sect.~\ref{tosources}).  
}

\subsection{Application to solar-like stars}
\label{solar-likes}

Sequences of Doppler images can be used to measure 
the surface differential rotation of solar-like stars and its time variability, as done by, e.g., 
\citet{Donatietal03} and \citet{Jeffersetal07} in the cases of \object{AB Dor} and \object{LQ Hya}. \citet{Lanza06} discussed the
implications of the observed changes of the surface differential rotation on the internal
dynamics of their convection zones, assuming that the angular velocity is
constant over cylindrical surfaces co-axial with the rotation axis. The present model
allows us to relax the Taylor-Proudman constraint on the internal angular velocity, but 
some different assumptions must be introduced to obtain the internal
torques in those active stars. Here we assume that the variation of the surface
differential rotation is entirely due to the Maxwell stresses of the internal magnetic
fields, localized in the overshoot layer below the convection zone, as in the interface dynamo model by, e.g., 
\citet{Parker93}. 
The interior model of the Sun can be applied also to \object{AB Dor} and \object{LQ Hya} because they
have a similar relative depth of the convection zone. Therefore, the basic quantities can be 
scaled according to the stellar parameters, as explained in \citet{Lanza06}.
Following \citet{Donatietal03}, we consider a surface differential rotation of the form:
\begin{equation}
\Omega(\mu, t) = \Omega_{\rm eq} (t) - d\Omega(t) \mu^{2}, 
\end{equation}
where $\Omega_{\rm eq}$ is the equatorial angular velocity and $d\Omega$ the latitudinal shear, both
regarded as time-dependent. Since $P_{0}^{(1,1)} = 1$ and $\mu^{2} = \frac{1}{5} +\frac{4}{15} P_{2}^{(1,1)}  $, 
it can be recast in the form of Eq.~(\ref{omegadevel}) leading to:
\begin{eqnarray}
\Omega_{\rm eq} - \frac{1}{5} d\Omega = \sum_{k} \alpha_{0k}(t) \zeta_{0k}(R), \nonumber \\
- \frac{4}{5} d\Omega = \sum_{k} \alpha_{2k}(t) \zeta_{2k}(R),
\label{omegastar}
\end{eqnarray}
where $R$ is the radius of the star. 
The functions $\beta_{nk}(t)$ can be computed by assuming that the timescale of variation of 
the differential rotation $t_{\rm DR}$ is significantly shorter than the timescales for angular momentum
transport, as given by $\lambda_{nk}^{-1}$. This assumption is 
justified in the case of rapidly rotating stars by the observed variation timescales of the order of a few years 
together with the expected rotational quenching of 
the viscosity \citep[see, e.g., ][]{Kichatinovetal94}.
 Therefore, Eq.~(\ref{alphaeq}) leads to $\alpha_{nk} (t) \simeq t_{\rm DR} \beta_{nk}(t)$.
The angular velocity can be written in a form similar to Eq.~(\ref{A_sol}) by introducing
 an appropriate Green function. For instance,
considering the first of Eqs.~(\ref{omegastar}), we find:
\begin{equation}
\Omega_{\rm eq} - \frac{1}{5} d\Omega  =  \frac{3}{8\pi} \int_{V} \left( \nabla \cdot {\vec \tau}_{1} \right) G_{\rm s}(R,r^{\prime}) d V^{\prime},
\end{equation}
where:
\begin{equation}
G_{\rm s} (r, r^{\prime}) = \sum_{k} \zeta_{0k} (r) \zeta_{0k} (r^{\prime}),   
\label{greenstar}  
\end{equation}
and the integration is extended over the stellar convection zone. 
Assuming that $\nabla \cdot {\vec \tau}_{1} $ is different from zero only in an overshoot layer of volume $V_{\rm o}$
and applying considerations similar to those of Sects.~\ref{sol_method} and \ref{loc_method},  
we find a lower limit for the variation of the averaged perturbation term: 
\begin{equation}
\langle \Delta (\nabla \cdot {\vec \tau}_{1}) \rangle_{\rm V_{\rm o}} 
\geq \left( \frac{8\pi}{3 t_{\rm DR}} \right) \left[ \frac{\Delta \Omega_{\rm eq} - \frac{1}{5} \Delta (d\Omega) }{V_{\rm o} M} 
\right], 
\label{stellar_est}
\end{equation}
where $\Delta \Omega_{\rm eq}$ and $\Delta (d\Omega)$ are the amplitudes of variation of the differential
rotation parameters, and $M$ is the maximum of $G_{\rm s} (R, r^{\prime})$ in the overshoot layer. 

This result made use of the limited information we can get from 
surface differential rotation. However, in the near future, 
the observations of the rotational splittings of  stellar oscillations promise to 
give information on the internal rotation 
and its possible time variations. Since only the 
modes of low degrees ($\ell \leq 3$) are detectable in disk-integrated measurements, the 
spatial resolution of the derived internal angular velocity profile is very low.
\citet{Lochardetal05} considered the case in which only the mean radial profile of 
$\Omega$ is measurable.   

In view of such an additional information accessible through asteroseismology, 
let us consider a more general case
in which some average of the internal angular velocity of the star, say $\omega_{\rm m}(t)$, can be 
measured as a function of the time: 
\begin{equation}
\omega_{\rm m}(t) = \int_{r_{\rm b}}^{R} \int_{-1}^{1} \omega(r, \mu ,t) w(r, \mu) dr d\mu,  
\label{omegam}
\end{equation} 
where $w$ is an appropriate weight function that takes into account the averaging effects of 
the limited spatial resolution, and $r_{\rm b}$ is the radius at the base of the stellar convective envelope.
 Let us introduce an auxiliary weight function:
\begin{equation}
w_{1}(r, \mu) \equiv \frac{w(r, \mu)}{(1-\mu^{2}) \rho r^{4}},
\end{equation}
which will not diverge toward the poles ($\mu = \pm 1$) and the 
surface ($\rho = 0$) because the weight function $w$ is localized into the stellar interior and
goes to zero rapidly enough toward the rotation axis and the
stellar surface. We can develop 
the function $w_{1}$ as:
\begin{equation}
w_{1} (r, \mu) = \sum_{n}^{N_{\rm w}} \sum_{k}^{K_{\rm w}} w_{nk} \zeta_{nk}(r) P_{n}^{(1,1)}(\mu),
\end{equation}
where the summation can be truncated at some low
orders, say, $N_{\rm w}$ and  $K_{\rm w}$,  
because  the function $w_{1}$ is not sharply localized in the stellar interior. 
Considering Eq.~(\ref{omegam}), we find:
\begin{equation}
\omega_{\rm m} (t) = \sum_{n}^{N_{\rm w}} \sum_{k}^{K_{\rm w}} \frac{1}{2 \pi F_{n}} w_{nk} \alpha_{nk}.
\label{omegamdevel}
\end{equation} 
For the sake of simplicity, we assume now that the time scale of variation of the functions $\alpha_{nk}(t)$ is
long with respect to $\lambda_{nk}^{-1}$ so that Eq.~(\ref{alphaeq}) gives: $\beta_{nk} \simeq \lambda_{nk} \alpha_{nk}$.
Considering Eq.~(\ref{betaeq2}), we find:
\begin{equation}
\omega_{\rm m}(t) = - \int_{V} {\vec \tau}_{1} \cdot \left\{ 
 \sum_{n}^{N_{\rm w}} \sum_{k}^{K_{\rm w}}  \frac{w_{nk}}{2 \pi \lambda_{nk}} \nabla \left[ \zeta_{nk} P_{n}^{(1,1)} \right]\right\}.  
\label{omegamfin}
\end{equation}
Equation~(\ref{omegamfin}) can be used to find a lower limit to the average of $| {\vec \tau}_{1}|$ over the convection 
zone volume. If we indicate by $M_{w}$ the maximum of the function: 
\begin{equation}
\left| \sum_{n}^{N_{\rm w}} \sum_{k}^{K_{\rm w}}  \frac{w_{nk}}{2 \pi \lambda_{nk}} \nabla \left[ \zeta_{nk} P_{n}^{(1,1)} \right] \right|
\end{equation}
over the volume of the convection zone, we find:
\begin{equation}
\langle | {\vec \tau}_{1}(t)| \rangle \equiv \frac{\int_{V} | {\vec \tau}_{1}| dV}{V} \geq \frac{|\omega_{\rm m}(t)|}{M_{\rm w} V}. 
\end{equation}

\section{Application to the Sun}

\subsection{Interior model, eigenvalues and eigenfunctions}
\label{interior_mod}

\begin{figure}
\psfig{file=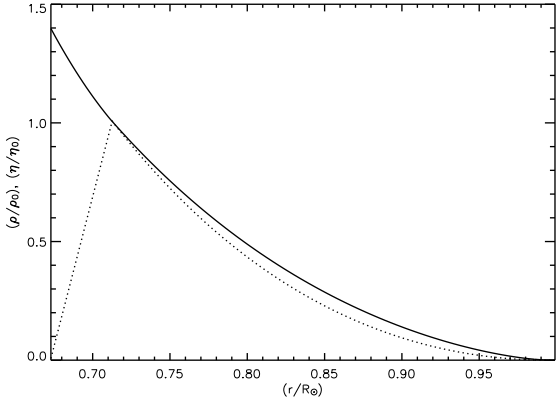,width=8cm}
\caption{
The ratio of the density to the density at the base of the convection zone  ($\rho  / \rho_{0} $ --
solid line) and the ratio of the turbulent dynamical viscosity to the turbulent viscosity at the
base of the convection zone ($ \eta_{t}  /\eta_{0}$ -- dotted) versus the
fractionary radius $r/R_{\odot}$ in our solar interior model. The density at the base of the convection zone
is $\rho_{0} = 0.1875 $ g cm$^{-3}$ and the turbulent viscosity is 
$ \eta_{0}  = 2.56 \times 10^{12}$ g cm$^{-1}$ s$^{-1}$, respectively. A linear increase of 
$\eta_{\rm t}$ between the relative radii 0.673 and 0.713 is assumed to  account for the effects of 
overshooting convection.
}
\label{fig1}
\end{figure}

A model of the solar interior can be used to specify the functions $\rho(r)$ and $\eta_{\rm t}(r)$ that
appear in our equations. While the density stratification can be determined with an accuracy better
than $0.5$\%, the turbulent dynamical viscosity is uncertain by at least one order of magnitude and it is estimated
from the mixing-length theory according to the formula:
\begin{equation}
\eta_{t} = \frac{1}{3} \alpha_{\rm ML} \rho u_{\rm c} H_{\rm p},
\label{etat}
\end{equation}
where  $\alpha_{\rm ML}$ is the ratio of the mixing-length to the pressure scale height $H_{\rm p}$ and 
$u_{\rm c}$ is the convective velocity
given by: 
\begin{equation}
u_{\rm c} = \left( \frac{\alpha_{\rm ML} L_{\odot} }{40 \pi r^{2} \rho}\right)^{\frac{1}{3}}, 
\label{vconv}
\end{equation}
where $L_{\odot}$ is the luminosity of the Sun. 
In our computations we adopt $\alpha_{\rm ML} = 1.5$ and assume the solar model S for the interior quantities\footnote{See
http://bigcat.ifa.au.dk/$\sim${jcd}/solar\_models/} \citep{Christensenetal96}. In that model, the base of the convection zone is
at $r = 0.713 R_{\odot}$. We consider also the effect of an overshoot layer extending between $r= 0.673 R_{\odot}$ and
the base of the convection zone within which the turbulent dynamical viscosity is 
assumed to increase linearly from zero up to the value
at the base of the convection zone. 
The density and the turbulent viscosity are
plotted in Fig.~\ref{fig1} where their values have been normalized at the values at the base of the convection zone, respectively. 
%%%%%%%%%%%%%%%%%%%%%%%%%%%%%%%%%%%%%%%
%%%%%%%%%%%%%%%%%%%%%%%%%%%%%%%%%%%%%%%%

The basic equations of our model (i.e., \ref{angvelp}, \ref{s1} and \ref{tau1}) can be made nondimensional
by adopting as the unit of length the solar radius $R_{\odot}$, as the unit of density $\rho_{0}$, i.e., the density at
the base of the solar convection zone, and as a unit of time $t_{0} = \rho_{0} R_{\odot}^{2}/\eta_{0}$, where
$\eta_{0}$ is the turbulent viscosity at the base of the convection zone. 
As a matter of fact, the value of the diffusion coefficients estimated from the mixing-length theory leads 
to a too short period for the solar cycle in mean-field dynamo models. \citet{Covasetal04} adopted a 
turbulent magnetic diffusivity  $ \nu_{\rm t} = 3 \times 10^{11}$ cm$^{2}$ s$^{-1}$ 
to get a sunspot cycle of $\sim 11$ yr. This implies 
$\eta_{0} \sim \rho \nu_{\rm t} = 5.62 \times 10^{10}$ g cm$^{-1}$ s$^{-1}$ in our model.

The  radial eigenfunctions $\zeta_{nk}$ and the Jacobian polynomials
$P_{n}^{(1,1)}$ have been computed from the respective Sturm-Liouville problem equations
by means of the Fortran 77 subroutine sleign2.f\footnote{http://www.math.niu.edu/SL2/} 
\citep{Baileyetal01}. For the radial eigenfunctions $\zeta_{nk}$, the Sturm-Lioville problem
has been solved with Neumann boundary conditions at both ends, set at $0.675 R_{\odot}$ and
$0.99 R_{\odot}$ to avoid  divergence at the surface. For the Jacobian polynomials,
limit point boundary conditions have been adopted at $\mu = \pm 1$.  

{ Note that, from a rigorous point of view, it would be better to use 
the helioseismic estimate of $\frac{\partial \Omega}{\partial r}$ at $r=0.99$ where we fixed our outer boundary for
the computation of the radial eigenfunctions instead of the stress-free boundary condition that is valid only at the
surface. However, the differences are confined to the outermost layer of the solar convection zone, 
the moment of inertia of which is so small that there are no pratical consequences.} 

The eigenvalues $\lambda_{nk}$ and the eigenfunctions $\zeta_{nk}$ computed by sleign2.f 
have been compared with those computed by means of the code introduced by \citet{Lanza06b}.
The relative differences in the eigenvalues and in the eigenfunctions are lower than 1.5\% for $n \leq 14$, $k \leq 10$. However,
 some problems of convergence of the numerical algorithm used by sleign2.f have been found for $ k\geq 20$, particularly 
for $n \geq 30$, so we decided to limit its application up to $k = 19$. 

The Jacobian polynomials and the eigenvalues computed by sleign2.f are very good up to $n =30$, as it has been found
by comparison with their analytic expressions up to $n=10$ and their asymptotic expressions for $n\geq 12$. 
We conclude that for $n \geq 30$ and $k \geq 20$ it is better to use the asymptotic formulae 
(\ref{zeta_asymp}) and (\ref{jacobi_asymp}) instead of the numerically computed $\zeta_{nk}$ and $P_{n}^{(1,1)}$.  

The eigenvalue $\lambda_{nk}$ gives the inverse of the characteristic timescale of 
angular momentum transfer of the mode corresponding to $\zeta_{nk}$ 
under the action of the turbulent viscosity. The longest timescale corresponds to
the lowest eigenvalue, i.e., $\lambda_{20}^{-1} = 0.078$ in nondimensional units. 
It corresponds to a time scale of 
$0.86$ yr with the turbulent viscosity given by the mixing-length theory 
and to $39.5$ yr with $\eta_{0}=5.62 \times 10^{10}$
g cm$^{-1}$ s$^{-1}$.  

%The successive eigenvalue is
%$\lambda_{01} = 132.06$ in dimensionless units.   

\subsection{Localization functions for the source of the torsional oscillations}
\label{loc_f_descr}

The available data on the torsional oscillations are displayed with a typical
radial resolution of $0.05$ $R_{\odot}$ and a latitudinal resolution of $15^{\circ}$ in 
\citet[][]{Howeetal05,Howeetal06}. The rotational inversion kernels of 
\citet{Schouetal98} show a higher radial resolution close to the surface, but, given the
small amplitude of the torsional oscillations, the choice of a uniform resolution of $0.05$ $R_{\odot}$
seems to be better. 

\begin{figure}
\psfig{file=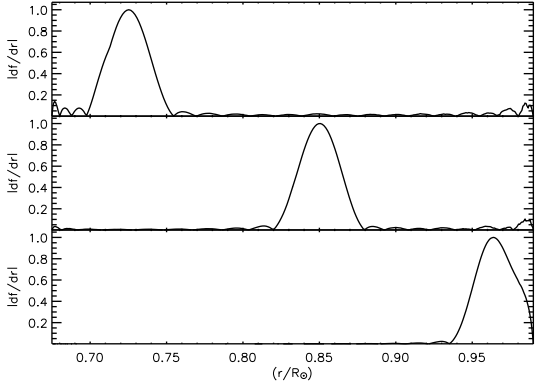,width=8cm}
\caption{
The modulus of the radial localization kernel versus the relative radius 
for the function (\ref{loc_funct}), as obtained by truncating the series of 
the radial eigenfunctions at $K_{\rm m }= 19$,
for three different intervals centered at $r=0.725$ (upper panel), $0.85$ 
(middle panel) and
$0.965$ $R_{\odot}$ (lower panel), respectively, all with an
amplitude of $r_{2} - r_{1} = 0.05$ $R_{\odot}$. The modulus of the derivative has been normalized to its maximum value. 
}
\label{fig2}
\end{figure}

We have found that the best results on the localization of the source term with the method
outlined in Sect.~\ref{loc_method} are obtained with localization functions that depend on
$r$ or  $\mu$ only and that have a smooth derivative. As a typical function to probe the
radial localization, we  adopt:
\begin{equation}
f(r)= \left\{ \begin{array}{ll}
                 0 & \mbox{for $r \leq r_{1}$,} \\
                 1 + \sin \left[ \pi \left( \frac{r -r_{1}}{r_{2} - r_{1}} \right)  -\frac{\pi}{2} \right] 
                   & \mbox{for $ r_{1} \leq r \leq r_{2}$,} \\
                 2 & \mbox{for $r \geq r_{2}$.}
              \end{array}
      \right.
\label{loc_funct}
\end{equation}
This function has  zero derivative, except in the interval $]r_{1}, r_{2}[$ where its derivative
varies smoothly reaching a maximum in the mid of the interval. In Fig.~\ref{fig2} we plot the 
modulus of the derivative $ \left| \frac{df}{dr} \right| $, as given by the sum of the series of the radial
eigenfunctions truncated at $k =19$, for three different intervals centered at $0.725$, $0.85$ and
$0.965$ $R_{\odot}$, respectively, all with an amplitude of $r_{2} - r_{1} = 0.05$ $R_{\odot}$. 
The derivative is well approximated by the truncated series, with small sidelobes
the amplitude of which increases toward the surface of the Sun because the eigenfunctions 
scale as $(\rho \eta_{\rm t})^{-\frac{1}{4}}$, according to the asymptotic expression (\ref{zeta_asymp}).
%%%%%%%%%%%%%%%%%%%%%%%%%%%%%%%%%%%%%%%
%%%%%%%%%%%%%%%%%%%%%%%%%%%%%%%%%%%%%%%%

\begin{figure}
\psfig{file=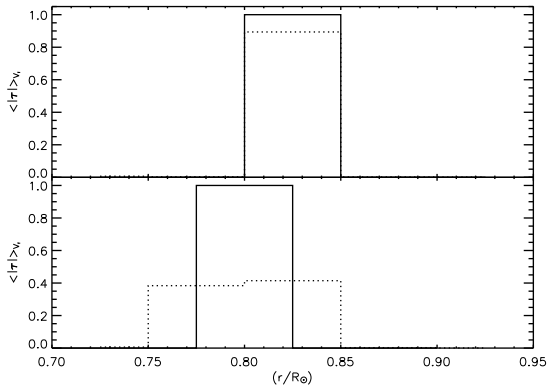,width=8cm}
\caption{
{\it Upper panel:} Test case of the application of the inversion method introduced in Sect.~\ref{loc_method}. An input profile  
with ${\vec \tau}_{1} = 1.0 \hat{\vec r}$  in nondimensional units between $0.80$ and $0.85$ $R_{\odot}$ (plotted
as the solid line) is used to simulate a noiseless profile of angular velocity perturbation.
The reconstructed lower-limit profile according to Eq.~(\ref{lower_tau1})
is plotted as a dotted line in the same panel. {\it Lower panel:} The same as in the upper panel, but with an
input profile localized between 0.775 and 0.825 $R_{\odot}$. 
}
\label{fig3}
\end{figure}

The performance of the localization method introduced in Sect.~\ref{loc_method} has been tested with
simulated data in the absence of noise. The results of two  such tests are plotted in Fig.~\ref{fig3}
where we consider  the case of a purely radial perturbation ${\vec \tau}_{1} = \tau_{1}\hat{\vec r}$ localized within
an interval of amplitude $0.05$ $R_{\odot}$.
Its time dependence is assumed to be purely cosinusoidal and the corresponding coefficients 
$\beta_{nk}$ are computed by means of Eq.~(\ref{betaeq2}) up to the orders $n=38$ and $k=19$. From the 
$\beta_{nk}$, the coefficients $\alpha_{nk}$ are computed by means of Eq.~(\ref{alphabeta}) and the simulated angular velocity
perturbation follows from Eq.~(\ref{omegadevel}).

 When the interval in which ${\vec \tau}_{1}$ is localized coincides with one of the inversion
intervals $[r_{1}, r_{2}]$, the lower limit for $| {\vec \tau}_{1}|$ turns out to be $\sim 80$\%$-90$\% of the value
assumed in the simulation (cf. Fig.~\ref{fig3} upper panel). The agreement increases up to $90$\%$-95$\% if we consider 
a case in which $\tau_{1r}$ has a constant sign and put the value of the derivative
$\frac{df}{dr}$ in the denominator of Eq.~(\ref{lower_tau1}) instead of its modulus. 
This happens because the modulus of the derivative 
increases the effects of the sidelobes by increasing the value of the denominator in Eq.~(\ref{lower_tau1}). 
When the interval in which the assumed ${\vec \tau}_{1}$ is localized does not coincide with an interval of 
the inversion grid, the inversion method still performs well distributing the contributions among neighbour intervals
nearly in the correct proportion (cf. Fig.~\ref{fig3}, lower panel).  
%%%%%%%%%%%%%%%%%%%%%%%%%%%%%%%%%%%%%%%
%%%%%%%%%%%%%%%%%%%%%%%%%%%%%%%%%%%%%%%%
The case of simulations including  a noise component is straightforward to treat thanks to the linear character of
our inversion method. The inverted value turns out to be the sum of the inverted noiseless value and of the contribution 
coming from the inversion of the noise. Their  amplitude ratio is equal to the ratio of the amplitudes of the input
noise to the input signal (for constant relative signal errors), as discussed in Sect.~\ref{loc_method}.

\begin{figure}
\psfig{file=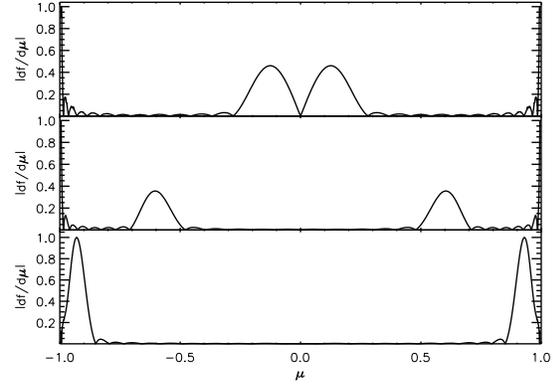,width=8cm}
\caption{
The modulus of the latitudinal localization kernel versus $\mu$ for the function (\ref{loc_funct_mu}), 
as obtained by truncating the series of the Jacobian polynomials at $N_{\rm m }= 30$, for three different 
intervals of $\mu$ in the Northern hemisphere, i.e., $[0, 0.26]$ (upper panel), $[0.5, 0.7]$  (middle panel) and
$[0.87, 0.99]$ (lower panel), respectively. Note the symmetry of the modulus of the kernel with respect to the    
equator. The value of $\left| \frac{df}{d\mu} \right|$
is normalized at its maximum at $\mu = \pm 1$. }
\label{fig4}
\end{figure}

The localization in latitude can be sampled by means of a localization function of the kind: 
\begin{equation}
f(\mu)= \left\{ \begin{array}{ll}
                 0 & \mbox{for $ -1 < \mu \leq -\mu_{2}$,} \\
                 1 + \sin \left[ \pi \left( \frac{\mu + \mu_{2}}{\mu_{2} - \mu_{1}} \right)  -\frac{\pi}{2} \right] 
                   & \mbox{for $ -\mu_{2} \leq \mu \leq -\mu_{1}$,} \\
                 2 & \mbox{for $-\mu_{1} \leq \mu \leq \mu_{1}$,} \\
                 1 + \sin \left[ \frac{\pi}{2} - \pi \left( \frac{\mu -\mu_{1}}{\mu_{2}- \mu_{1}} \right) \right]
                   & \mbox{for $\mu_{1} \leq \mu \leq \mu_{2}$,} \\
                 0 & \mbox{for $\mu_{2} \leq \mu < 1 $.} 
              \end{array}
      \right.
\label{loc_funct_mu}
\end{equation}
Such a function is symmetric with respect to the equator so that its derivative $\frac{df}{d\mu}$ is antisymmetric, 
as it is the source term ${\vec \tau}_{1}$ leading to a symmetric perturbation of the angular velocity. 
The localization function has a derivative always equal to zero except in two intervals $]-\mu_{2}, -\mu_{1}[$ and
$]\mu_{1}, \mu_{2}[$, symmetric with respect to the solar equator. Its representation by means of the Jacobian
polynomials with degree up to $N_{\rm} = 30$ is given in Fig.~\ref{fig4}. Note that the maximum of $\left| \frac{df}{d\mu} \right|$
is reached in two sharp peaks at $\mu = \pm 1$. However, they are so narrow that their contribution to the integral in 
Eq.~(\ref{lower_tau1}) is modest in comparison to those of the broader peaks in the intervals $[-\mu_{2}, -\mu_{1}]$ 
and $[\mu_{1}, \mu_{2}]$.   
%%%%%%%%%%%%%%%%%%%%%%%%%%%%%%%%%%%%%%%
%%%%%%%%%%%%%%%%%%%%%%%%%%%%%%%%%%%%%%%%
Several tests have been performed, as in the case of the radial localization, to assess the performance of the proposed method.
The results are similar to those obtained for the radial case and are not discussed here. 

We conclude that our choice of $N_{\rm m} = 30$ and 
$K_{\rm m} =19$ is perfectly adequate to invert the available data on the solar torsional oscillations to derive information on
the location of the perturbation term within the convection zone. 

\subsection{Results on the sources of the torsional oscillations}
\label{tosources}

The data plotted in Fig.~4 of \citet{Howeetal06} can be used for an illustrative application of the inversion
methods introduced in Sect.~\ref{loc_method}. They are given with a sampling of $15^{\circ}$
between the equator and $60^{\circ}$ of 
latitude, i.e., in the range in which the rotational inversion techniques perform better \citep[cf. ][]{Schouetal98}. 
The features distinguishable on the plots indicate an actual radial resolution of $\sim 0.05$ $R_{\odot}$, 
in agreement with the sampling adopted in Fig.~3 of \citet{Howeetal05}. 
An average statistical error of about 30\% can be assumed for the amplitude, whereas the phase errors become
very large below $0.80$ $R_{\odot}$, especially  at low latitudes, because of the uncertainty in the reconstruction of the signal
in the deep layers. Note that the error intervals reported in Fig.~4 of \citet{Howeetal06} indicate only how the 
particular method solution (here an OLA inversion of SoHO/MDI data) would vary with a different realization of the
input data affected by a randon Gaussian noise. Unfortunately, they do not give the statistical ranges 
in which the true values of the amplitude and phase are likely to lie. This is not a major limitation in the 
context of the present study because we  aim at illustrating the capabilities of the proposed approach rather than 
derive definitive conclusions. 

To perform our inversion, we interpolate linearly the values of the amplitude and phase over the grid used to compute the
radial eigenfunctions and the Jacobian polynomials. We assume that the amplitude is 
zero at the poles and increases linearly toward $60^{\circ}$ of latitude whereas the phase is constant poleward of
$60^{\circ}$ of latitude. 

\begin{figure}
\psfig{file=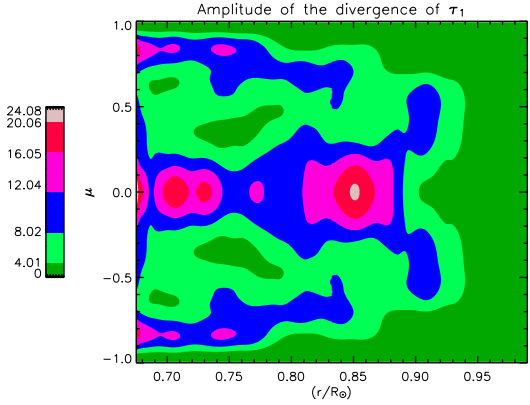,width=8cm}
\caption{
The isocontours of the function $A_{\tau} (r, \mu)$ as defined by Eq.~(\ref{atau}) in the case of $\eta_{0}=2.56 \times 10^{12}$ 
g cm$^{-1}$ s$^{-1}$. The  scale on the left indicates the ranges corresponding to the different colors and
 is in units of $10^{5}$ g cm$^{-1}$ s$^{-2}$. The relative statistical uncertainty of $A_{\tau}$ is of $\sim 30$\%, as it follows
from the relative uncertainty of the data.}
\label{div_amp}
\end{figure}

\begin{figure}
\psfig{file=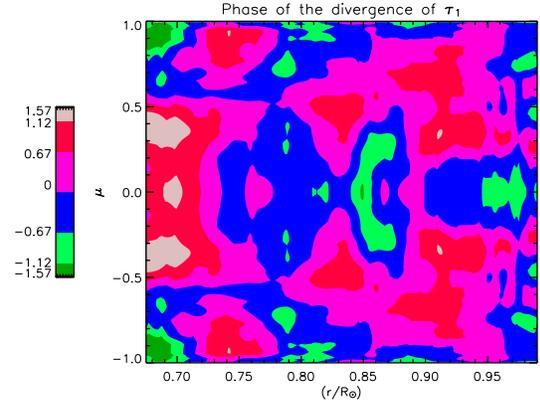,width=8cm}
\caption{
The isocontours of  $\Phi_{\tau} (r, \mu)$ as defined by Eq.~(\ref{phitau}) in the case of $\eta_{0}=2.56 \times 10^{12}$ 
g cm$^{-1}$ s$^{-1}$. The phase ranges from $-\frac{\pi}{2}$ to $\frac{\pi}{2}$. The scale on the left indicates the
phase ranges corresponding to the different colors and is in radians. }
\label{div_phase}
\end{figure}

\begin{figure}
\psfig{file=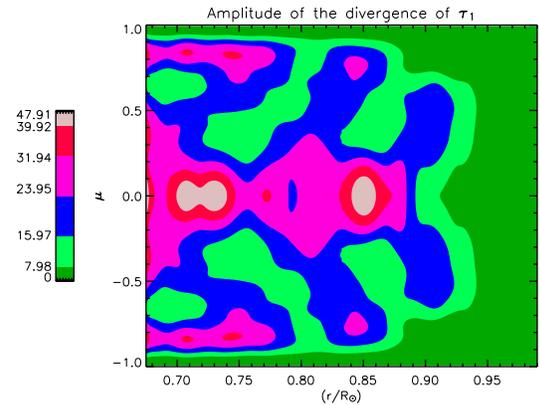,width=8cm}
\caption{
The same as Fig.~\ref{div_amp} in the case of $\eta_{0}=5.62 \times 10^{10}$ 
g cm$^{-1}$ s$^{-1}$. The scale on the left is in units of $10^{3}$ g cm$^{-1}$ s$^{-2}$. The relative statistical
uncertainty of $A_{\tau}$ is of about 30\%.}
\label{div_amp_red}
\end{figure}

\begin{figure}
\psfig{file=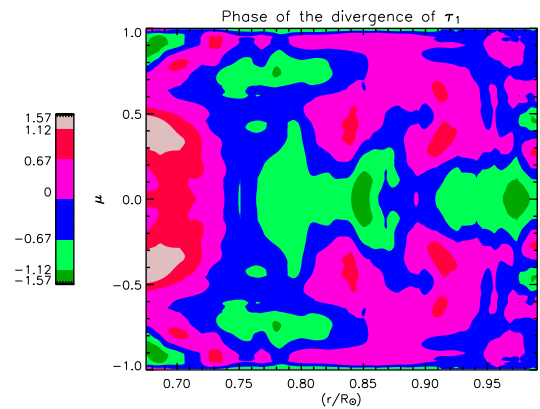,width=8cm}
\caption{
The same as Fig.~\ref{div_phase} in the case of $\eta_{0}=5.62 \times 10^{10}$ 
g cm$^{-1}$ s$^{-1}$. }
\label{div_phase_red}
\end{figure}

The divergence of the angular momentum flux perturbation ${\vec \tau}_{1}$ can be obtained from Eq. (\ref{div_loc}).
We define its amplitude and phase as:
\begin{equation}
A_{\tau} \equiv \sqrt{[\nabla \cdot {\vec \tau}_{1}^{(c)}]^{2} + [\nabla \cdot {\vec \tau}_{1}^{(s)}]^{2} },
\label{atau}
\end{equation}
\begin{equation}
\Phi_{\tau} \equiv \tan^{-1} \left( \frac{\nabla \cdot {\vec \tau}_{1}^{(c)}}{\nabla \cdot {\vec \tau}_{1}^{(s)}} \right).
\label{phitau}
\end{equation}
They are plotted in Figs.~\ref{div_amp} and \ref{div_phase}, respectively, in the case of $\eta_{0}=2.56 \times 10^{12}$ 
g cm$^{-1}$ s$^{-1}$. A suitable smoothing has been applied
to have a resolution of $\sim 0.05$ $R_{\odot}$ in the radial coordinate and  $\approx 15^{\circ}$ in latitude.
 %%%%%%%%%%%%%%%%%%%%%%%%%%%%%%%%%%%%%%%
%%%%%%%%%%%%%%%%%%%%%%%%%%%%%%%%%%%%%%%%
 %%%%%%%%%%%%%%%%%%%%%%%%%%%%%%%%%%%%%%%
%%%%%%%%%%%%%%%%%%%%%%%%%%%%%%%%%%%%%%%%
In order to show the effect of a smaller value of the turbulent viscosity, we plot in Figs.~\ref{div_amp_red} and 
\ref{div_phase_red} the isocontours of $A_{\tau}$ and $\Phi_{\tau}$ for $\eta_{\rm t} = 5.62 \times 10^{10}$ 
g cm$^{-1}$ s$^{-1}$. 
 %%%%%%%%%%%%%%%%%%%%%%%%%%%%%%%%%%%%%%%
%%%%%%%%%%%%%%%%%%%%%%%%%%%%%%%%%%%%%%%%
 %%%%%%%%%%%%%%%%%%%%%%%%%%%%%%%%%%%%%%%
%%%%%%%%%%%%%%%%%%%%%%%%%%%%%%%%%%%%%%%%
The amplitude of the perturbation term is higher close to the equator because the data in Fig.~4 of 
\citet{Howeetal06} mainly sample the low-latitude branch of the torsional oscillations. The relative 
maxima of $A_{\tau}$ are reached close to the base of the convection zone and at a radius of $\sim 0.85$
$R_{\odot}$ and their locations show only  a minor dependence on the value of $\eta_{0}$. Conversely, the value of $\eta_{0}$ significantly 
affects $\Phi_{\tau}$, as it follows from  
Eq.~(\ref{bcoeff}). When $\eta_{0}$ is large, the timescales for angular momentum exchange, as given by 
the inverse of the lowest eigenvalues, are significantly shorter than the eleven-year cycle and 
the perturbations are almost in phase with the angular velocity variations. When 
$\eta_{0}$ is sufficiently low, the timescales for angular momentum exchange 
become  comparable or longer than the eleven-year cycle so the torsional oscillations lag behind the perturbations.
Considering Fig.~\ref{div_phase}, we see that the phase is in agreement with that in 
Fig.~4 of \citet{Howeetal06}, except for $r \gtsim 0.93$ $R_{\odot}$. The disagreement in those layers
is due to the small values of  $\nabla \cdot {\vec \tau}_{1}$ close to the surface that makes the
corresponding phases uncertain (cf. Fig.~\ref{div_amp}). The phase lag is apparent by comparing Fig.~\ref{div_phase_red}
with Fig.~\ref{div_phase}, especially close to the equator for $ 0.72 \leq r/R_{\odot} \leq 0.92$. It is less evident
near the base of the convection zone and in the upper layers due to the smaller values of $\nabla \cdot {\vec \tau}_{1}$ in those
regions.

{
It is interesting to note that the dependence of the amplitude and phase of the torsional oscillations on the turbulent
viscosity can lead to its estimate in the framework of mean-field models
\citep[see, e. g., ][ for details]{Rudigeretal86,Rudiger89}. Specifically, \citet{Rempel07} finds that 
a mean turbulent kinematic viscosity about one order of magnitude smaller than the mixing-length estimate is needed  
to reproduce the polar branch of the torsional oscillations.  
}

\begin{figure}
\psfig{file=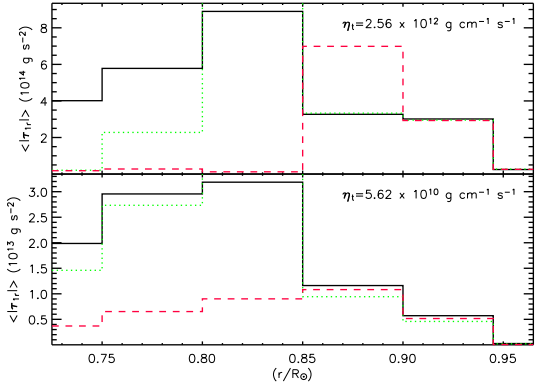,width=8cm}
\caption{{\it Upper panel:}
The lower limit of the amplitude of the perturbation term $|{\vec \tau}_{\rm 1r}|$ averaged over spherical shells of 
thickness $0.05$ $R_{\odot}$ versus the relative radius; different linestyles and colors refer to the depth at which
the torsional oscillations are assumed to vanish: black solid line -- oscillations extending down to the base of the convection zone;
green dotted  line -- oscillations extending down to 0.80 $R_{\odot}$; red dashed line -- oscillations extending down to 0.90
$R_{\odot}$. Results are obtained assuming a turbulent dynamical viscosity at the base of the convection zone
$\eta_{0} = 2.56 \times 10^{12}$ g cm$^{-1}$ s$^{-1}$. {\it Lower panel:} As in the upper panel, but for  
$\eta_{0} = 5.62 \times 10^{10}$ g cm$^{-1}$ s$^{-1}$. The relative statistical uncertainties are in all the cases of
about 30\%, as it follows from the uncertainties of the data. 
 }
\label{rad_loc}
\end{figure}

\begin{figure}
\psfig{file=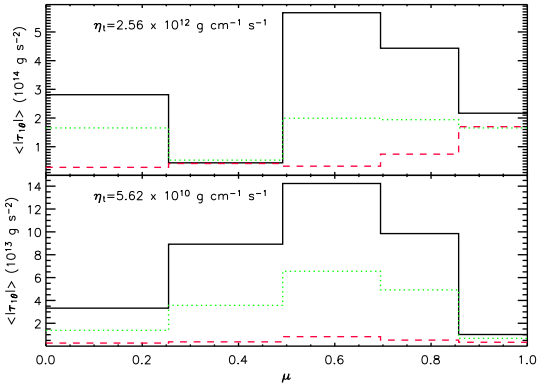,width=8cm}
\caption{{\it Upper panel:}
The lower limit of the amplitude of the perturbation term $|{\vec \tau}_{1\theta}|$ averaged over different latitude zones 
versus their average value of $\mu$. Different linestyles are used to indicate the results for different penetration depths 
of the torsional oscillations, as in Fig.~\ref{rad_loc}. Results are obtained assuming 
a turbulent dynamical viscosity at the base of the convection zone $\eta_{0} = 2.56 \times 10^{12}$ g cm$^{-1}$ s$^{-1}$. 
{\it Lower panel:} As in the upper panel, but for  
$\eta_{0} = 5.62 \times 10^{10}$ g cm$^{-1}$ s$^{-1}$. The relative statistical uncertainties are in all the cases of 
about 30\%, as it follows from the uncertainties of the data. 
 }
\label{mu_loc}
\end{figure}

Lower limits for the modulus of the angular momentum flux vector $|{\vec \tau}_{1}|$ can be derived 
from Eq.~(\ref{lower_tau1}). From the lower limits on $|{\vec \tau}_{1}^{(c)}|$ and $|{\vec \tau}_{1}^{(s)}|$, 
a lower limit on $|{\vec \tau}_{1}|= \sqrt{[{\vec \tau}_{1}^{(c)}]^{2} + [{\vec \tau}_{1}^{(s)}]^{2}}$ can be derived 
and it is plotted in Figs.~\ref{rad_loc} and \ref{mu_loc} for the radial and latitudinal 
localizations described in Sect.~\ref{loc_f_descr}, respectively.  
Given the uncertainty in the penetration depth of the torsional 
oscillations, in addition to the results for  a penetration down to the base of the convection zone,
we plot also those for penetration depths of $0.80$ and $0.90$ $R_{\odot}$, respectively. They are obtained simply 
by assuming that the oscillation amplitude has the value in Fig.~4 of
\citet{Howeetal06} above the penetration depth and drops to zero immediately below it.
 
When the oscillations penetrate down to the base of the convection zone, the maximum of the perturbation term
is reached between $0.80$ and $0.85$ $R_{\odot}$. It is mainly 
localized into two latitude zones, i.e., within $\pm 15^{\circ}$ from the 
equator and  between $30^{\circ}$ and $60^{\circ}$. Note that our data refer mainly to the low-latitude branch of the oscillations,
so the possible source at latitude $> 60^{\circ}$ cannot be detected. When the turbulent dynamical viscosity is reduced with 
respect to the mixing-length value, the amplitude of the perturbation term drops, but its radial and latitudinal localizations
are not greatly affected, except for a shift of the nearly equatorial band towards higher latitudes. 

\begin{figure}
\psfig{file=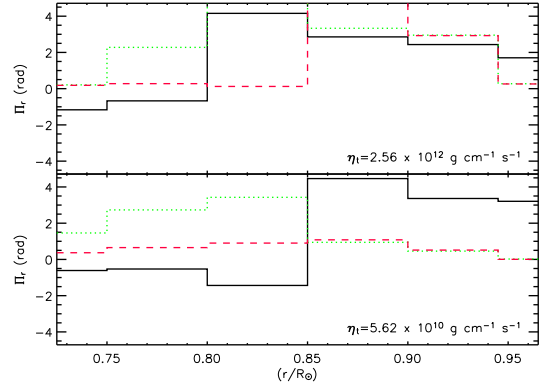,width=8cm}
\caption{{\it Upper panel:}
The phase lag $\Pi_{\rm r}$ between $B_{\rm r}$ and $B_{\phi}$ as derived by Eqs.~(\ref{brbphi}) averaged over spherical shells of 
thickness $0.05$ $R_{\odot}$ versus the relative radius.  
Different linestyles are used to indicate the results for different penetration depths 
of the torsional oscillations, as in Fig.~\ref{rad_loc}.
Results are obtained assuming a turbulent dynamical viscosity at the base of the convection zone
$\eta_{0} = 2.56 \times 10^{12}$ g cm$^{-1}$ s$^{-1}$. {\it Lower panel:} As in the upper panel, but for  
$\eta_{0} = 5.62 \times 10^{10}$ g cm$^{-1}$ s$^{-1}$. 
 }
\label{fig11}
\end{figure}

\begin{figure}
\psfig{file=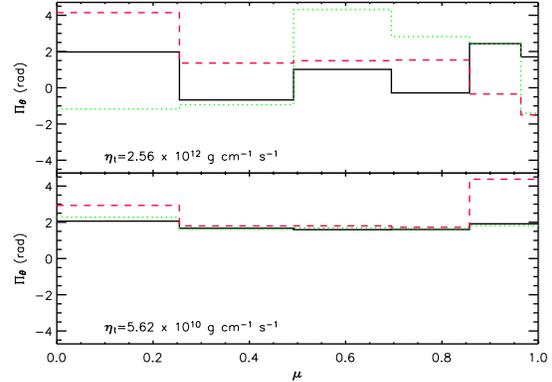,width=8cm}
\caption{{\it Upper panel:}
The phase lag $\Pi_{\theta}$ between $B_{\theta}$ and $B_{\phi}$ as derived by Eqs.~(\ref{brbphi}) averaged 
over different latitude zones 
versus their average value of $\mu$. 
Different linestyles are used to indicate the results for different penetration depths 
of the torsional oscillations, as in Fig.~\ref{rad_loc}.
 Results are obtained assuming a turbulent dynamical viscosity at the base of the convection zone
$\eta_{0} = 2.56 \times 10^{12}$ g cm$^{-1}$ s$^{-1}$. {\it Lower panel:} As in the upper panel, but for  
$\eta_{0} = 5.62 \times 10^{10}$ g cm$^{-1}$ s$^{-1}$.  
 }
\label{fig12}
\end{figure}

{ Under the hypothesis that the Lorentz force is the only source of the torsional oscillations, 
the amplitude of the Maxwell stress $B_{\rm r}B_{\phi}$ can be estimated from the lower limit of $|{\vec \tau}_{\rm 1r}|$ and 
it turns out to be $\sim 2.7 \times 10^{5}$ G$^{2}$ in the case of $\eta_{0} = 2.56 \times 10^{12}$ g cm$^{-1}$ s$^{-1}$.
 Considering that the poloidal field $B_{\rm r}$ is of the order of $1-10$ G, this
leads to very high toroidal fields in the bulk of the convection zone, that would be highly unstable because of 
magnetic buoyancy. On the other hand, with $\eta_{0} = 5.62 \times 10^{10}$ g cm$^{-1}$ s$^{-1}$, the Maxwell stress is 
$B_{\rm r}B_{\phi} \simeq 8.1 \times 10^{3}$ G$^{2}$ which leads to a toroidal field intensity of the order of 
$10^{3}$ G. It may be stably stored for timescales comparable to the solar cycle thanks to the effects of the
downward turbulent pumping in the convection zone \citep[cf., ][ and references therein]{Brandenburg05}.

Note that the maximum radial stress $| B_{\rm r}B_{\phi}|$ is reduced by a factor of $\sim 30$ when $\eta_{\rm t}$ is decreased 
from $2.56 \times 10^{12}$ to $5.62 \times 10^{10}$ g cm$^{-1}$ s$^{-1}$, whereas the maximum of $| B_{\theta} B_{\phi}|$ is 
reduced only by a factor of $\sim 4$ (cf. Figs.~\ref{rad_loc} and \ref{mu_loc}). This mainly reflects the predominance
of the radial gradient over the latitudinal gradient of the angular velocity perturbation in the deeper layers of the 
solar convection zone. 

The average phase lags $\Pi_{\rm r}$ and $\Pi_{\theta}$ between $B_{\rm r}$ and $B_{\theta}$ and the azimuthal field
$B_{\phi}$, as derived by the method in Sect.~\ref{meanfieldeff}, are plotted in Figs.~\ref{fig11} and \ref{fig12}, respectively.
It is interesting to note that $B_{\rm r} B_{\phi} >0$ below $\sim 0.85$ $R_{\odot}$ when 
$\eta_{\rm t} = 5.62 \times 10^{10}$ g cm$^{-1}$ s$^{-1}$, in agreement with the finding of most dynamo models in which the
azimuthal field is produced by the stretching of the radial field in the low-latitude region,
 where $\frac{\partial \Omega}{\partial r} > 0$ for $r < 0.95$ $R_{\odot}$
\citep[cf., e.g., ][]{RudigerHollerbach04,SchlichenStix95}.
Above $\sim 0.85$ $R_{\odot}$, the phase relationship between the radial and the azimuthal fields leads to 
$B_{\rm r} B_{\phi} < 0$ with a phase lag of $\sim \pi$, 
in agreement with the early finding that the photospheric zone equatorward of the activity belt is rotating
faster than that at higher latitudes \citep[cf. ][]{Rudiger89}. Note that $\frac{\partial \Omega}{\partial r} $ becomes negative 
for $r > 0.95$ $R_{\odot}$ at low latitudes, leading to a reversal of the phase relationship between 
the two field components in the outer layers of the solar convection zone. Our limited spatial resolution and the uncertainty
of the measurements of the torsional oscillations inside the Sun might explain why we find the transition from a mostly positive
to a negative $B_{\rm r} B_{\phi}$ at $\sim 0.85$ $R_{\odot}$. 

The phase lag between $B_{\theta}$ and $B_{\phi}$ depends remarkably 
on the colatitude for $\eta_{\rm t} = 2.56 \times 10^{12}$ g cm$^{-1}$ s$^{-1}$, whereas it is almost constant for
$\eta_{\rm t} = 2.56 \times 10^{12}$ g cm$^{-1}$ s$^{-1}$, leading in the latter case mostly to 
$B_{\theta} B_{\phi} < 0$. This, together with $\frac{\partial \Omega}{\partial \theta} > 0$,  suggests 
that the toroidal field is mainly produced by the stretching of the radial field. 
}

On the other hand, if the angular momentum transport leading to the torsional oscillations is 
produced only by a perturbation of the meridional flow, as in the thermal wind model, then a very small perturbation follows
from the lower limit of $|{\vec \tau}_{1}|$. For $\eta_{0} = 2.56 \times 10^{12}$ g cm$^{-1}$ s$^{-1}$,
we find a minimum amplitude of the meridional flow component oscillating with the eleven-year cycle of $\approx 3$ cm s$^{-1}$ at a
depth of $0.85$ $R_{\odot}$ and a latitude of $45^{\circ}$. Note, however, that 
if we estimate $|{\vec \tau}_{1}|$ from its divergence and the typical lengthscale of its variations in Fig.~\ref{div_amp},
we find a value about one order of magnitude larger, that is in agreement with the estimate by \citet{Rempel07}. A similar argument
applies also to the estimate of the Maxwell stresses made above.

The average  kinetic energy variation associated with the
torsional oscillations, as computed from Eq.~(\ref{kvar}), is $\langle {\cal T}\rangle _{\rm c} = 4.88\times 10^{28}$ erg, 
whereas the average dissipated power is $ 6.7\times 10^{26}$ erg s$^{-1}$, 
when $\eta_{0} = 2.56 \times 10^{12}$ g cm$^{-1}$ s$^{-1}$, and only 
$1.5 \times 10^{25}$ erg s$^{-1}$, when $\eta_{0} = 5.6 \times 10^{10}$ g cm$^{-1}$ s$^{-1}$. 

When the  torsional oscillations are assumed to be confined to shallower and shallower layers, 
the location of the perturbation term is shifted closer and closer to the surface and it 
becomes more and more uniformly distributed in latitude. 
Its amplitude shows a remarkable decrease because of the smaller moment of inertia of the surface layers.  
 %%%%%%%%%%%%%%%%%%%%%%%%%%%%%%%%%%%%%%%
%%%%%%%%%%%%%%%%%%%%%%%%%%%%%%%%%%%%%%%%
 %%%%%%%%%%%%%%%%%%%%%%%%%%%%%%%%%%%%%%%
%%%%%%%%%%%%%%%%%%%%%%%%%%%%%%%%%%%%%%%%
%%%%%%%%%%%%%%%%%%%%%%%%%%%%%%%%%%%%%%%
%%%%%%%%%%%%%%%%%%%%%%%%%%%%%%%%%%%%%%%%
%%%%%%%%%%%%%%%%%%%%%%%%%%%%%%%%%%%%%%%
%%%%%%%%%%%%%%%%%%%%%%%%%%%%%%%%%%%%%%%%

\section{Application to solar-like stars}

The results of Sect.~\ref{solar-likes} can be applied to the variation of the surface differential rotation
observed in, e.g., \object{LQ Hya} between 1996.99 and 2000.96 \citep[see Table 2 of ][]{Donatietal03}. Assuming an internal
structure analogous to the solar one and that the differential rotation variations observed at the surface are
representative of those at a depth of $0.99$ $R$, we can estimate a lower limit for $| \nabla \cdot {\vec \tau}_{1}|$ in the 
overshoot layer from Eq.~(\ref{stellar_est}).
 It is used to estimate the Maxwell stresses by assuming that: 
$ r_{\rm o} \frac{B_{\rm p}B_{\phi}}{\tilde{\mu}} \simeq \delta r | \nabla \cdot {\vec \tau}_{1}|$, where
$r_{\rm o} = 0.67$ $R$ is the radius of the overshoot layer and $\delta r=0.04$ $R$ its thickness. In such a way, 
the minimum magnetic field
strength turns out to be: $B_{\rm min} = \sqrt{B_{\rm p}B_{\phi} } \sim 2500$ G. However, if we assume a poloidal field strength of
$\sim 100$ G, as indicated by the Zeeman Doppler imaging, we find an azimuthal field of $B_{\phi} \sim 6.2 \times 10^{4}$ G,
which is in the range of the values estimated by \citet{Lanza06} in the framework of the Taylor-Proudman hypothesis. Note that the
result is independent of $\eta_{0}$ in the limit $t_{\rm DR} \ll \lambda_{nk}^{-1}$. 
Although this application is purely illustrative, it suggests that our method can be applied to derive estimates of the
internal magnetic torques (as well as other sources of angular momentum transport) when asteroseismic results will
become available, in combination with surface rotation measurements, to further constrain  rotation variations in solar-like stars. 

\section{Discussion and conclusions}

We introduced a general solution of the angular momentum
transport equation that takes into account the density stratification and the radial dependence of the turbulent
viscosity $\eta_{\rm t}$. Its main limitation is due to the uncertainty of the 
turbulent viscosity in stellar convection zones.  It is interesting to note that the method of the separation of 
variables to solve Eq.~(\ref{angmomeq}) can be applied also when $\eta_{\rm t}$ is the product of a function of the radius
by one of the latitude. However, when $\eta_{\rm t}$ depends on the latitude,
the angular eigenfunctions are no longer Jacobian polynomials. 

Our formalism can be applied to compute the response of a turbulent convection zone to prescribed time-dependent
Lorentz force and  meridional circulation. From the 
mathematical point of view, it is a generalization of that of \citet{Rudigeretal86} and can be easily 
compared with it by considering that: $P_{n}^{(1,1)} (\mu) = \frac{2}{n+2} \frac{dP_{n+1}}{d\mu} = -  \frac{2 \sin \theta}{n+2}
P_{n}^{1}(\theta)$, where $P_{n}$ is the Legendre polynomial of degree $n$ and $P_{n}^{1} = -\frac{dP_{n}}{d\theta}$. 
Moreover, our method can be used to estimate the torques leading to the  angular momentum redistribution
within the solar (or stellar) convection zone, thus generalizing the approach suggested by \citet{Kommetal03}.
The main limitation, in addition to the uncertainty of the turbulent viscosity, comes from the low
resolution and the limited accuracy of the present data on  solar torsional oscillations, particularly in the deeper layers
of the convection zone. Actually, those layers are the most important because  torsional oscillations with an amplitude of 
$\sim 0.5$\%$-1$\% of the solar angular velocity 
extending down to the base of the convection zone lead to Maxwell stresses with an intensity of at least $\approx 
8 \times 10^{3}$ G$^{2}$ around 0.85 $R_{\odot}$ or a perturbation of the order of several percents of the meridional flow speed
 at the same depth \citep[cf., e.g., ][]{Rempel07}. 
{
If the torsional oscillations are due solely to the Maxwell stresses associated with the mean field of the solar dynamo,
we can estimate also the phase relationships between  $B_{\rm r}$, $B_{\theta}$ and $B_{\phi}$. Our preliminary results
indicate that $B_{\rm r}B_{\phi} >0$ in the layers below $\approx 0.85$ $R_{\odot}$ and $B_{\rm r}B_{\phi} <0$ in the outer
layers which, together with helioseismic measurements of the internal angular velocity, suggest that the toroidal field is
mainly produced by the stretching of the poloidal field by the radial shear.
}

Future helioseismic measurements may improve our knowledge of the 
torsional oscillations, essentially by extending the time series of the data or by means of space-borne 
instruments, like those foreseen for the Solar Dynamic Observatory \citep[see, e.g., ][]{Howeetal06}. 
On the other hand, asteroseismic measurements may open the possibility of investigating similar phenomena in solar-like stars,
particularly in those young, rapidly rotating objects showing 
variations of the angular velocity one or two orders of magnitude larger than the Sun. 

\begin{acknowledgements}
The author wishes to thank an anonymous Referee for valuable comments and Professor G.~R\"udiger for interesting discussion. 
Solar physics and active star research at INAF-Catania Astrophysical Observatory and the Department of Physics
and Astronomy of Catania University is funded by MIUR ({\it Ministero dell'Universit\`a e della Ricerca}), and by {\it Regione Siciliana}, whose financial support is gratefully
acknowledged.

This research has made use of the ADS-CDS databases, operated at the CDS, Strasbourg, France.
\end{acknowledgements}

\appendix

\section{Proof of convergence of the  Green function series}

The convergence of the Green function series in Eqs.~(\ref{green}) can be studied by considering 
the asymptotic formulae for $\zeta_{nk}$ and the 
Jacobian polynomials, i.e., for $n \gg 1$ and $k \gg 1$. In the asymptotic limit, those series can be
written as: 
\begin{eqnarray}
\lefteqn{\sum_{n} \sum_{k}  \frac{F_{n}}{\lambda_{nk}} \zeta_{nk}(r) \zeta_{nk}(r^{\prime}) P_{n}^{(1,1)} (\mu) 
 P_{n}^{(1,1)} (\mu^{\prime}) = } \nonumber & & \\
 & & {\sum_{n} h_{n} \, \sqrt{F_{n}} P_{n}^{(1,1)} (\mu)},
\label{pseries}
\end{eqnarray}
where:
\begin{equation}
h_{n} \equiv \sum_{k} \frac{\sqrt{F_{n}}}{\lambda_{nk}} \zeta_{nk}(r) \zeta_{nk}(r^{\prime}) P_{n}^{(1,1)} (\mu^{\prime}).  
\label{hseries}
\end{equation}

From the asymptotic formulae it follows that
for a point in the
domain $[r_{\rm b}, R_{\odot}] \times [r_{\rm b}, R_{\odot}] \times ]-1, 1[ $, the
quantity $\left| \sqrt{F_{n}} \zeta_{nk}(r) \zeta_{nk}(r^{\prime})  P_{n}^{(1,1)} (\mu^{\prime}) \right| $ is 
limited with an upper bound independent of $n$ and $k$. Therefore, the series (\ref{hseries}) that define the coefficients 
$h_{n}$ are uniformly convergent in that
domain if the series $\sum_{k} \frac{1}{\lambda_{nk}}$ converges. 

This can be proven by considering the 
inequalities (\ref{eigen_ineq}) and the formula:
\begin{equation}
\sum_{k=1}^{\infty} \frac{z}{a k^{2} + y} = \frac{1}{2} \left( \frac{z}{y} \right) \left[ g\left( \frac{y}{a}\right) 
\sqrt{\frac{y}{a}} - 1 \right],
\label{simplefrac}
\end{equation}
where $z$, $a \not= 0$ and $y$ are real numbers and the function $g(x)$ is defined as:
\begin{equation}
g(x) \equiv \pi \left[ \frac{\exp(\pi x) + \exp(-\pi x)}{\exp(\pi x) - \exp(-\pi x)} \right]. 
\end{equation}
Eq.~(\ref{simplefrac}) follows from the equality (1.217.1) of  \citet{GradshteynRizhik94}. In this way,
we find:
\begin{eqnarray}
\lefteqn{\frac{m}{2n(n+3)Q}\left\{ g\left[ \frac{n(n+3) Q l^{2}}{\pi^{2} P}\right]  
\sqrt{\frac{n(n+3) Q l^{2}}{\pi^{2}P}}  -1 \right\} \leq } \nonumber \\
& & \leq \sum_{k=1}^{\infty} \frac{1}{\lambda_{nk}} \leq  \\
\leq 
& & \frac{M}{2n(n+3)q}\left\{ g\left[ \frac{n(n+3) q l^{2}}{\pi^{2} p}\right]  
\sqrt{\frac{n(n+3) q l^{2}}{\pi^{2}p}}  -1 \right\}. \nonumber  
\label{kseries} 
\end{eqnarray}
This result indicates that $h_{n} \sim \frac{1}{\sqrt{n(n+3)}}$ for $n \gg 1$. 
Since the eigenfunctions $\sqrt{F_{n}} P_{n}^{(1,1)}(\mu)$ form an orthonormal set, 
the convergence of the
series in~(\ref{pseries}) follows by  the Riesz-Fisher Theorem given that the
series $\sum_{n} h_{n}^{2} \sim \sum_{n} \frac{1}{n(n+3)}$ converges. 
In such a way, the uniform convergence of the series in Eqs.~(\ref{green}) in the domain
$[r_{\rm b}, R_{\odot}] \times [r_{\rm b}, R_{\odot}] \times ]-1, 1[ \times ]-1, 1[ $ is proven.

\end{document}